\documentclass[11pt,a4paper]{scrartcl}

\usepackage{ILD}
\usepackage{arydshln}

\newdate{date}{16}{04}{2021}
\date{\displaydate{date}}
\ildphyspub{2021}{001}

\usepackage[symbol]{footmisc}
\usepackage{graphicx}   
\usepackage{color}
\usepackage{enumitem}
\usepackage{lineno}
\usepackage{mathtools}
\usepackage{amsmath}
\usepackage{verbatim}

\usepackage{array}
\newcolumntype{L}[1]{>{\raggedright\let\newline\\\arraybackslash\hspace{0pt}}m{#1}}
\newcolumntype{C}[1]{>{\centering\let\newline\\\arraybackslash\hspace{0pt}}m{#1}}
\newcolumntype{R}[1]{>{\raggedleft\let\newline\\\arraybackslash\hspace{0pt}}m{#1}}

\newcommand{\beq}{\begin{equation}}
\newcommand{\eeq}{\end{equation}}
\def\bea{\begin{eqnarray}}
\def\eea{\end{eqnarray}}
\newcommand{\bei}{\begin{itemize}}
\newcommand{\eei}{\end{itemize}}

\def\={\,=\,}
\def\+{\,+\,}
\def\-{\,-\,}
\def\TeV{\ifmmode {\mathrm{Te\kern -0.1em V}}\else
                   \textrm{Te\kern -0.1em V}\fi\,}%
\def\GeV{\ifmmode {\mathrm{Ge\kern -0.1em V}}\else
                   \textrm{Ge\kern -0.1em V}\fi\,}%
\def\MeV{\ifmmode {\mathrm{Me\kern -0.1em V}}\else
                   \textrm{Me\kern -0.1em V}\fi\,}%
\def\keV{\ifmmode {\mathrm{ke\kern -0.1em V}}\else
                   \textrm{ke\kern -0.1em V}\fi\,}%
\def\eV{\ifmmode  {\mathrm{e\kern -0.1em V}}\else
                   \textrm{e\kern -0.1em V}\fi\,}%

\let\gev=\GeV

\def\iab{\mbox{ab$^{-1}$}}
\def\ifb{\mbox{fb$^{-1}$}}

\def\Rb{\ensuremath{R_{3}^{b}}\xspace}
\def\Rl{\ensuremath{R_{3}^{\ell}}\xspace}
\def\Rbl{\ensuremath{R_{3}^{b\ell}}\xspace}
\def\Rq{\ensuremath{R_{3}^{q}}\xspace}
\def\yc{\ensuremath{y_{cut}}\xspace}

\def\Kgamma{\ensuremath{E_{\gamma}}\xspace}
\def\Kreco{\ensuremath{K_{reco}}\xspace}

\def\epsilonb{\ensuremath{\epsilon_{b}}\xspace}
\def\epsilonc{\ensuremath{\epsilon_{c}}\xspace}
\def\epsilonuds{\ensuremath{\epsilon_{uds}}\xspace}
\def\epsilonb2{\ensuremath{\epsilon^{2}_{b}}\xspace}
\def\epsilonc2{\ensuremath{\epsilon^{2}_{c}}\xspace}
\def\epsilonuds2{\ensuremath{\epsilon^{2}_{uds}}\xspace}

\def\costheta{\ensuremath{\cos \theta}\xspace}

\def\bbbar{\ensuremath{b}\ensuremath{\overline{b}}\xspace}
\def\qqbar{\ensuremath{q}\ensuremath{\overline{q}}\xspace}
\def\ccbar{\ensuremath{c}\ensuremath{\overline{c}}\xspace}

\def\eeqq{\ensuremath{e^{\mbox{\scriptsize -}}}\ensuremath{e^{\mbox{\scriptsize +}}}\ensuremath{\rightarrow}\ensuremath{q}\ensuremath{\overline{q}}\xspace}
\def\ee{\ensuremath{e^{\mbox{\scriptsize -}}}\ensuremath{e^{\mbox{\scriptsize +}}}\xspace}

\def\eLpR{\ensuremath{e_L^{\mbox{\scriptsize -}}}\ensuremath{e_R^{\mbox{\scriptsize +}}}\xspace}
\def\eRpL{\ensuremath{e_R^{\mbox{\scriptsize -}}}\ensuremath{e_L^{\mbox{\scriptsize +}}}\xspace}

\def\bquark{\ensuremath{b}-quark\xspace}
\def\lquark{\ensuremath{\ell}-quark\xspace}
\def\lquarks{\ensuremath{\ell}-quarks\xspace}
\def\bquarks{\ensuremath{b}-quark\xspace}

\def\cquark{\ensuremath{c}-quark\xspace}

\def\Zboson{\ensuremath{Z}-boson\xspace}
\def\Zpole{\ensuremath{Z}-pole\xspace}

\def\pmp{\ensuremath{+-}\xspace}
\def\mpp{\ensuremath{-+}\xspace}

\def\MSbar{\ensuremath{\overline{MS}}\xspace}

\graphicspath{ {./logos/}{./figures/} }

\title{Prospects for the measurement of the \bquark mass at the ILC}
\addauthor{Juan Fuster}{\institute{1}} 
\addauthor{Adrian Irles}{\institute{1}} 
\addauthor{Germ\'an Rodrigo}{\institute{1}} 
\addauthor{Seidai Tairafune}{\institute{2}} 
\addauthor{Marcel Vos}{\institute{1}} 
\addauthor{Hitoshi Yamamoto}{\institute{1}\institute{2}} 
\addauthor{Ryo Yonamine}{\institute{2}} 

\addinstitute{1}{IFIC, Universitat de Val\`encia and CSIC, c./ Catedr\'atico Jos\'e Beltr\'an 2, E-46980 Paterna, Spain}
\addinstitute{2}{Tohoku University, 6-3, Aramaki Aza-Aoba, Aoba-ku, Sendai 980-8578, Japan}

\abstract{
This note presents an analysis of the potential of future high-energy electron-positron colliders to measure the \bquark mass.  We perform a full-simulation study of the measurement of the ratio of the three-jet rates in events with $b\bar{b}(g)$ and $q\bar{q}(g)$ production, \Rbl, and assess the dominant uncertainties, including theory and experimental systematic uncertainties. We find that the ILC "Higgs factory" stage, with an integrated luminosity of 2~\iab at $\sqrt{s}=$ 250~\gev can measure the \bquark \MSbar mass at a scale of 250~\gev ($m_b(250~\gev)$) with a precision of 1~\gev. From this result we extrapolate the potential of the GigaZ run
running at $\sqrt{s}= m_Z$. We expect $m_b(m_Z)$ can be determined with an 0.12~\gev{} uncertainty, exceeding the precision of the LEP and SLD measurements by a factor $\sim$3. }
\titlecomment{This work was carried out in the framework of the ILD concept group}

\begin{document}
\titlepage
\tableofcontents

\section{Introduction}

Quark masses in the Standard Model (SM) of particle physics are parameters of high 
importance and they are required to be determined experimentally with the highest
possible accuracy. In this document we will discuss the prospects of the \bquark mass 
measurement in future electron and positron colliders. 
In the past, several groups have extracted the bottom-quark mass from low energy data. The most precise recent extractions~\cite{Kiyo:2015ufa,Penin:2014zaa} rely on measurements of the mass spectrum of bottomonium bound states (i.e. the mass of the $\Upsilon(1S)$ resonance is known very precisely: $m_{\Upsilon(1S)} = 9.46030(26)~\gev$), perturbative QCD calculations, QCD sum rules and lattice. Other approaches use measurements of the $e^+e^- \rightarrow $ hadrons cross section as experimental input. These low-scale measurements dominate the PDG world average~\cite{Zyla:2020zbs} for the bottom-quark \MSbar mass $m_{b}(m_{b})$: 
\begin{equation}
  m_b(m_b) = 4.18^{+0.03}_{-0.02}~\gev,
\end{equation}

\noindent where $m_{b}(m_{b})$ stands for the \bquark mass at the scale 
$m_{b}$ in the \MSbar renormalization scheme. As parameters of the QCD 
Lagrangian, the quark masses in the \MSbar scheme are not constant, and vary 
with the energy scale of the process. This scale dependence is described 
with the renormalization group equations (RGE). Precise \bquark mass 
measurements at energy scales well above the quark mass itself in 
high-energy electron-positron colliders are of great interest as an evidence of 
the mass running as predicted by QCD. Motivated by the remarkable sensitivity of jet observables to the 
quark masses, a method to extract the bottom-quark mass from $Z$-pole data was proposed in 
Ref.~\cite{Bilenky:1994ad}. Three independent groups completed the necessary next-to-leading order (NLO) 
theoretical calculation of the three-jet rate for massive 
quarks~\cite{Rodrigo:1997gy,Bilenky:1998nk,Rodrigo:1999qg,Bernreuther:1997jn,Brandenburg:1997pu,Nason:1997tz,Nason:1997nw} that were successfully used to measure the \bquark mass far above threshold.

The three-jet fraction $R_3^{flav}$ is defined as follows:
\begin{equation}
    R_3^{flav} = \frac{\sigma^{3jet}_{flav} (y_{cut})}{\sigma_{flav}},
\end{equation}
where $\sigma^{3jet}_{flav}$ represents the cross section of $q\bar{q}+X$ -- being $flav$ the $q$-flavour,
$X$ the extra radiation (\emph{i.e.} a hard gluon) -- clustered as three separated jets as a function of the jet resolutions parameter $y_{cut}$ for a given jet clustering algorithm (e.g., the Durham~\cite{Catani:1991hj} and Cambridge~\cite{Dokshitzer:1997in} algorithms). The total inclusive cross section
of $q\bar{q}$ production is represented by $\sigma_{flav}$. When operating at the Z-pole, the three-jet fraction  $R_3^{flav}$
is expressed as:
\begin{equation}
    R_3^{flav} = \frac{\Gamma^{3jet}_{flav} (y_{cut})}{\Gamma_{flav}},
\end{equation}
where $\Gamma^{3jet}_{flav}$ is the width of the Z decaying to three jets with two quarks of flavour $flav$ in the final state and $\Gamma_{flav}$ is the total width of the Z decay involving two quarks of flavour $flav$.

In practice, the double ratio
\begin{equation}
  R_3^{b\ell} = \frac{R_3^b}{R_3^\ell} 
  \label{eq:r3bl}
\end{equation}
of the three-jet fractions for $b$-tagged events and $\ell$-tagged events ($\ell=u,d,s$) is used to determine $m_b(m_Z)$. It retains the excellent sensitivity to the bottom quark mass of $R_3^b$, while important systematic uncertainties cancel to some extent in the double ratio.

The first measurement of this type was performed by the DELPHI collaboration~\cite{Abreu:1997ey,Rodrigo:1997gy} using the LEP $Z$-pole data. Similar measurements were also performed with SLD data~\cite{Abe:1998kr,Brandenburg:1999nb} and by the ALEPH~\cite{Barate:2000ab} and OPAL~\cite{Abbiendi:2001tw} collaborations. DELPHI later improved its measurement from the three-jet rate in the Cambridge jet algorithm~\cite{Abdallah:2005cv} and added a measurement based on the four-jet rate~\cite{Abdallah:2008ac}. The values obtained for $m_b(m_Z)$ are summarized in Table~\ref{tab:mbmz}.

\begin{table}[ht]
\caption{\label{tab:mbmz} Measurements of the bottom-quark \MSbar mass at the scale $\mu=m_{Z}$, from three and four-jet rates with bottom quarks in $e^+e^-$ collisions at the $Z$-pole at LEP and SLD.}
\begin{tabular}{l|c|c}
experiment & $m_b(m_Z)$ [\gev] & comment \\ \hline
DELPHI     &  2.67 $\pm$ 0.25 (stat.) $\pm$ 0.34 (frag.) $\pm$ 0.27 (th.)               &     Ref.~\cite{Abreu:1997ey}    \\
SLD        &  2.56 $\pm$ 0.27 (stat.) $^{+0.28}_{-0.38}$ (syst.) $^{+0.49}_{-1.48}$ (th.)                &     Ref.~\cite{Abe:1998kr}, mass in Ref.~\cite{Brandenburg:1999nb} \\
ALEPH      & 3.27 $\pm$ 0.22 (stat.) $\pm$ 0.22 (exp.) $\pm$ 0.38 (had.) $\pm$ 0.16 (th.) & Ref.~\cite{Barate:2000ab} \\
OPAL      & 2.67 $\pm$ 0.03 (stat.) $^{+0.29}_{-0.37}$ (syst.) $\pm$ 0.19 (th.)  &    Ref.~\cite{Abbiendi:2001tw} \\ 
DELPHI     & 2.85 $\pm$ 0.18 (stat.) $\pm$ 0.13 (exp.) $\pm$ 0.19 (had.) $\pm$ 0.12 (th.) & Ref.~\cite{Abdallah:2005cv}, with Cambridge~\cite{Bilenky:1998nk} \\
DELPHI     & 3.76 $\pm$ 0.32 (stat.) $\pm$ 0.17 (syst.) $\pm$ 0.22 (had.) $\pm$ 0.90 (th.) & LO four-jet rate~\cite{Abdallah:2008ac} \\
\end{tabular}
\end{table}

Since then, the scale evolution of quark masses has been studied for the charm quark at HERA~\cite{Gizhko:2017fiu} and for the top quark at the LHC~\cite{Sirunyan:2019jyn} (see Ref.~\cite{Catani:2020tko} for a critical discussion of this measurement). The sensitivity of a future electron-positron collider to the running of the top quark mass is assessed in Ref.~\cite{Boronat:2019cgt}.

At a new electron-positron collider operated at the optimum energy for the Higgs-strahlung process (a "Higgs factory" in the jargon of the field) the bottom quark mass can be determined at yet higher scales. In analogy to the LEP/SLC measurements, jet rates in $e^+e^- \rightarrow b\bar{b} X$ production at $\sqrt{s}=$ 250~\gev can be used to extract the bottom quark mass. The measurement of $m_b(250~\gev)$ may enhance the significance of the evidence for the "running" of the bottom quark mass and probes the QCD evolution of the bottom quark mass out to much higher scales. At this higher scale, subtle effects from new massive states are expected to be enhanced~\cite{Bora:2012tx,PhysRevD.94.035016}. 

In this note, we assess the potential of a Higgs factory to measure $m_b(250~\gev)$. The precision of the measurement is estimated on the basis of a Monte Carlo simulation study of the $e^+e^- \rightarrow b\bar{b}$ process and the main background processes. These 
processes are generated at leading order and matched with a parton shower and hadronization generator. Predictions for the three-jet ratio \Rbl, its uncertainty and the sensitivity to the bottom quark mass are obtained with the NLO QCD calculation of Ref.~\cite{Bilenky:1998nk,Rodrigo:1999qg}. The detector response of the ILD experiment~\cite{ILD:2020qve} at the ILC~\cite{Bambade:2019fyw,Behnke:2013xla}\footnote{All results shown in this note are produced with the samples and ILCSoft (https://github.com/iLCSoft) version v01-16-p10.} is simulated with a detailed model of the experiment implemented in GEANT4~\cite{Agostinelli:2002hh}. The statistical uncertainty is then estimated taking into account the acceptance and selection efficiency, in a realistic scenario of the ILC~\cite{Barklow:2015tja} operation.   Experimental systematic uncertainties are determined with an in-situ method, while modelling and hadronization uncertainties are extrapolated from the LEP measurements. 

Operating a new electron-positron collider at the $Z$-pole, as is envisaged in the "GigaZ" program of the ILC and the "TeraZ" runs of circular collider projects, will enable a new measurement of $m_b(m_Z)$. With a much larger integrated luminosity, progress in theory and Monte Carlo event generators and much improved vertex detectors closer to the interaction point we would expect the experimental uncertainties to be reduced.
The detailed projection for the systematic uncertainties developed for the 250~\gev run is therefore used to assess the potential of a new  measurement of $m_b(m_Z)$. We extrapolate the most important uncertainties from our full-simulation study at 250~\gev to derive a rough estimate of the precision that can be achieved with the "GigaZ" program.

\section{Theory: mass effects in jet rates at NLO QCD}
\label{sec:theory}

In general, and with the exception of the top quark, the mass effects of the quarks are expected to be very small for inclusive observables (inclusive production cross section, etc) at high energies, since, by dimensional analysis, they are suppressed as the square of the ratio of the quark mass to the centre of mass energy. For example, for the \bquark this ratio is $m_b^2/m_Z^2 \simeq 10^{-3}$ at the $Z$-pole, and the suppression factor will be more evident at even higher energies, e.g., at the ILC Higgs factory stage \begin{equation}
    \frac{m_b^2}{(250~\gev)^2} \sim 10^{-4}~.
\end{equation}
Nevertheless, when more exclusive observables than the total cross section are considered, like jet cross sections, mass effects are enhanced as $(m_b^2/s) \log{\yc}$, where \yc is the resolution parameter that defines the jet multiplicity, then offering a unique opportunity to probe and measure the \bquark mass at high energies. 

Moreover, since quarks are not free particles, their masses can be considered as another coupling, and one has the freedom to use different quark mass definitions, e.g., the long-distance perturbative pole mass $M_b$ or the running mass $m_b(\mu)$ in the \MSbar scheme at a specific renormalization scale $\mu$, which is more suitable at short distances. At a fixed order in perturbation theory there is a residual dependence on which mass definition is used, as well as on the renormalization scale. The inclusion of higher orders to reduce these two uncertainties, due to mass definition and renormalization scale, is mandatory for an accurate description of the mass effects.

We recall the theory predictions~\cite{Rodrigo:1997gy,Bilenky:1998nk,Rodrigo:1999qg} 
for the observable in \eqref{eq:r3bl}, which admit the following perturbative expansion at next-to-leading order (NLO)
\begin{equation}
\Rbl = 1 + \frac{\alpha_S(\mu)}{\pi} a_0(\yc) + r_b \left(
b_0(r_b,\yc) + \frac{\alpha_S(\mu)}{\pi} b_1(r_b,\yc) \right)~,
\label{eq:r3pole}
\end{equation}
where $\ell = \{ u, d, s\}$ refers to the sum over the three light flavours. 
The function $a_0$ originates from triangle diagrams~\cite{Hagiwara:1990dx}. 
It is numerically very small ($a_0(0.01) \sim 0.04$ for both the Durham and Cambridge jet-clustering algorithms) and almost independent of the \bquark mass. The $b_0$ and $b_1$ functions give the leading order (LO) and NLO mass corrections, respectively, once the leading dependence on $r_b = M^2_b/s$, where $M_b$ is the \bquark pole mass, has been factorised out.

Using the known relationship between the pole mass and the \MSbar 
running mass,
\begin{equation}
M_b^2 = m_b^2 (\mu) \left[1 + \frac{2 \alpha_S(\mu)}{\pi}
\left( \frac{4}{3} - \log \frac{m_b^2(\mu)}{\mu^2}\right) \right]~,   
\end{equation}
we can re-express \eqref{eq:r3pole} in terms of the running mass $m_b(\mu)$. Then, keeping only terms of order ${\cal O}(\alpha_S)$ we obtain
\begin{equation}
\Rbl = 1 + \frac{\alpha_S(\mu)}{\pi} a_0(\yc) + \overline{r}_b(\mu) \left(
b_0(\overline{r}_b,\yc) + \frac{\alpha_S(\mu)}{\pi} \overline{b}_1(\overline{r}_b,\yc,\mu) \right)~,
\label{eq:r3running}
\end{equation}
where $\overline{r}_b(\mu) = m_b^2(\mu)/s$ and $\overline{b}_1(\overline{r}_b,\yc,\mu) = b_1(\overline{r}_b,\yc) + 2 b_0(\overline{r}_b,\yc)(4/3 - \log \overline{r}_b + \log(\mu^2/s))$. Although both expressions in \eqref{eq:r3pole} and \eqref{eq:r3running} are equivalent at the perturbative level, they give different answers since different higher order contributions have been neglected in each of them. The spread of the results gives an estimate of the size of missing higher order contributions, and thus of the theoretical uncertainty.

We extract the functions $a_0$, $b_0$, $b_1$ and $\overline{b}_1$ for the Durham and Cambridge jet-clustering algorithms from \cite{Rodrigo:1997gy,Bilenky:1998nk,Rodrigo:1999qg}. The Cambridge algorithm~\cite{Dokshitzer:1997in} reduces the formation of spurious jets formed with low transverse momentum particles that appear in the Durham algorithm at low \yc. Therefore, compared to Durham, the Cambridge allows to test smaller values of \yc while still keeping higher order corrections relatively small. This makes 
the calculation with the Cambridge algorithm more sensitive than with the Durham algorithm.

\begin{figure}[ht]
\centering
\includegraphics[scale=.75]{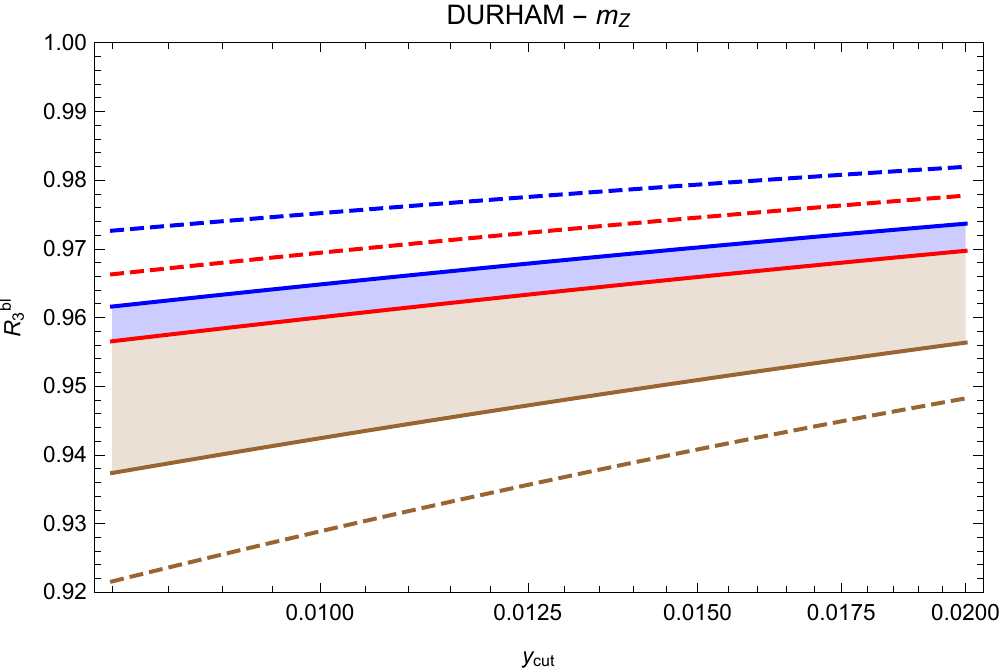}
\includegraphics[scale=.75]{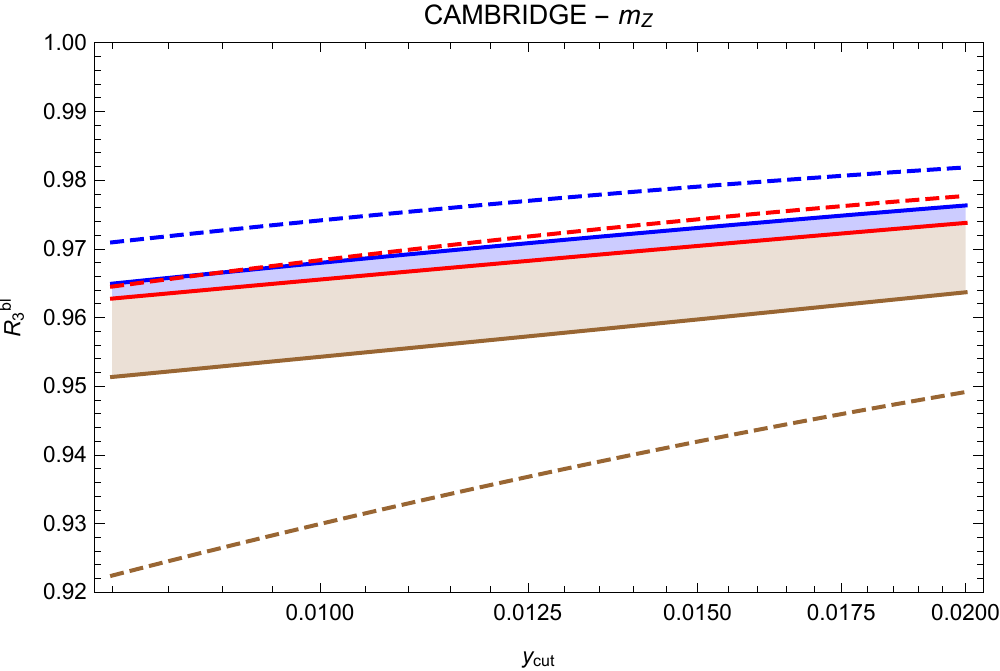}
\includegraphics[scale=.75]{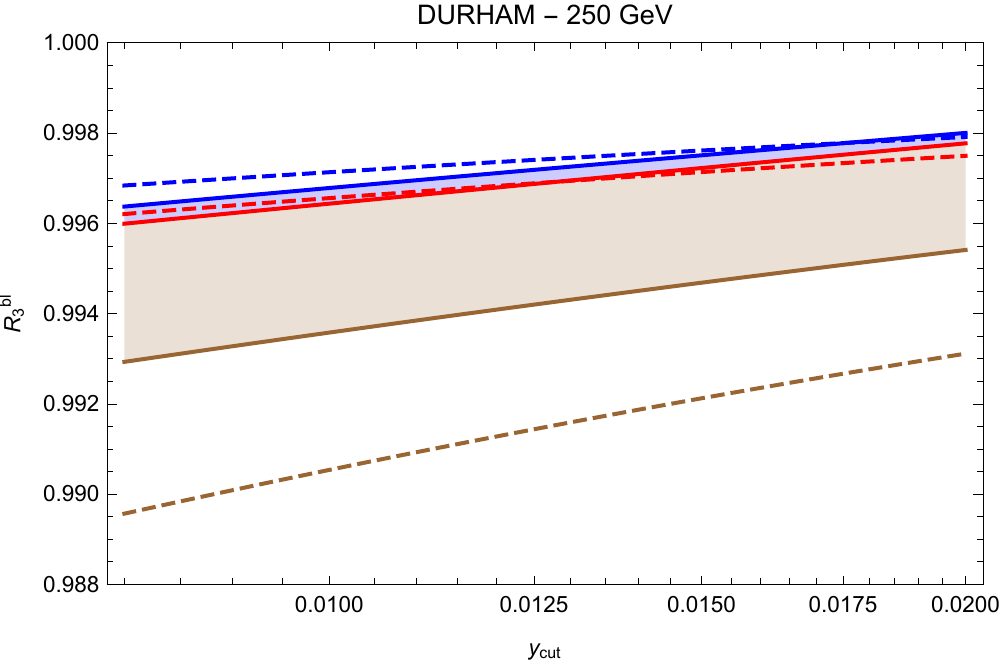}
\includegraphics[scale=.75]{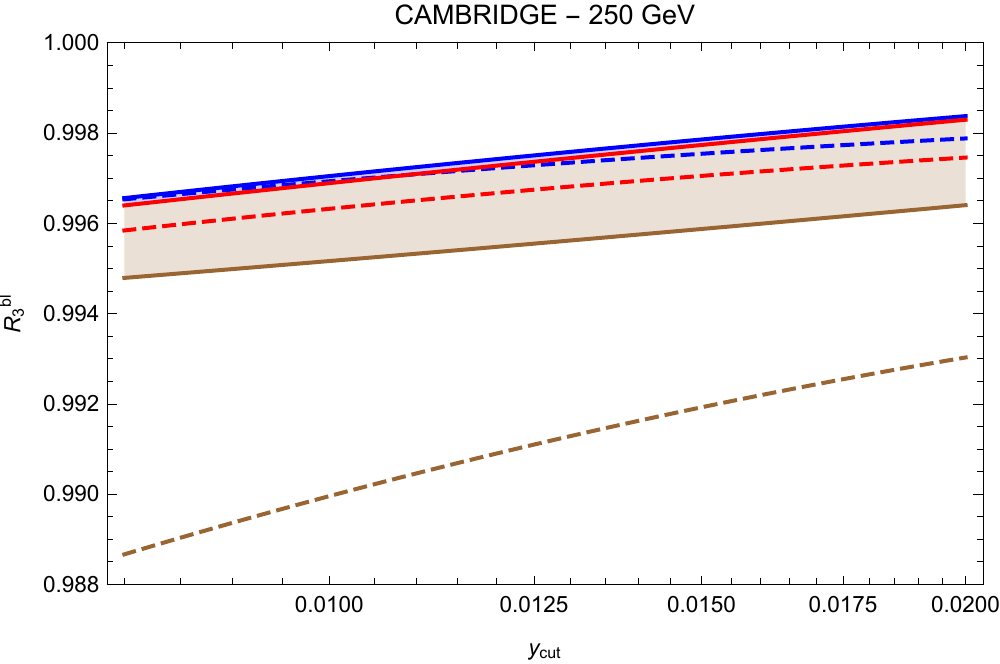}
\caption{Theory predictions for \Rbl in the Durham (left plots) and Cambridge (right plots) jet-clustering algorithm at the $Z$-pole scale (upper plots) and $\sqrt{s} = 250~\gev$ (lower plots). The dashed lines are LO, and the solid lines are NLO. The brown lines correspond to the theory predictions in terms of the pole mass and $\mu=\sqrt{s}$. The blue and red lines represent the theory predictions with the running mass and renormalization scales at $\mu = 2\sqrt{s}$ and $\mu = \sqrt{s}/2$, respectively. The theory uncertainty is estimated from the spread of the results, and is given by the shadowed band at NLO. } 
\label{fig:theoryr3bl}
\end{figure}

Our reference input numerical values for the \bquark running and pole masses are taken from the world averages of the 2020 PDG~\cite{Zyla:2020zbs}:
\begin{equation}
m_b(m_b) = 4.18^{+0.03}_{-0.02}~\gev~, \qquad 
M_b = 4.78 (6)~\gev~,
\label{eq:inputmb}
\end{equation}
together with the corresponding values for the strong coupling, and the masses of the $Z$ and Higgs bosons
\begin{equation}
\alpha_{S}(m_Z) = 0.1179 (10)~, \qquad
m_Z = 91.1876 (21)~\gev~, \qquad m_H = 125.10 (14)~\gev~.
\label{eq:input}
\end{equation}
We use the analytic solutions to the two-loop renormalization group equations in QCD from Refs.~\cite{Rodrigo:1997zd,Rodrigo:1993hc} to evolve the \bquark mass in Eq.~\eqref{eq:inputmb} from low energies to higher energies, 
\begin{equation}
  m_b(m_Z) = 2.97 (4)~\gev~,   
\end{equation}
and then to the ILC 250~\gev scale in the SM
\begin{equation}
  m_b(250~\gev) = 2.75 (4)~\gev~.   
\end{equation}

\begin{figure}[ht]
\centering
\includegraphics[scale=.75]{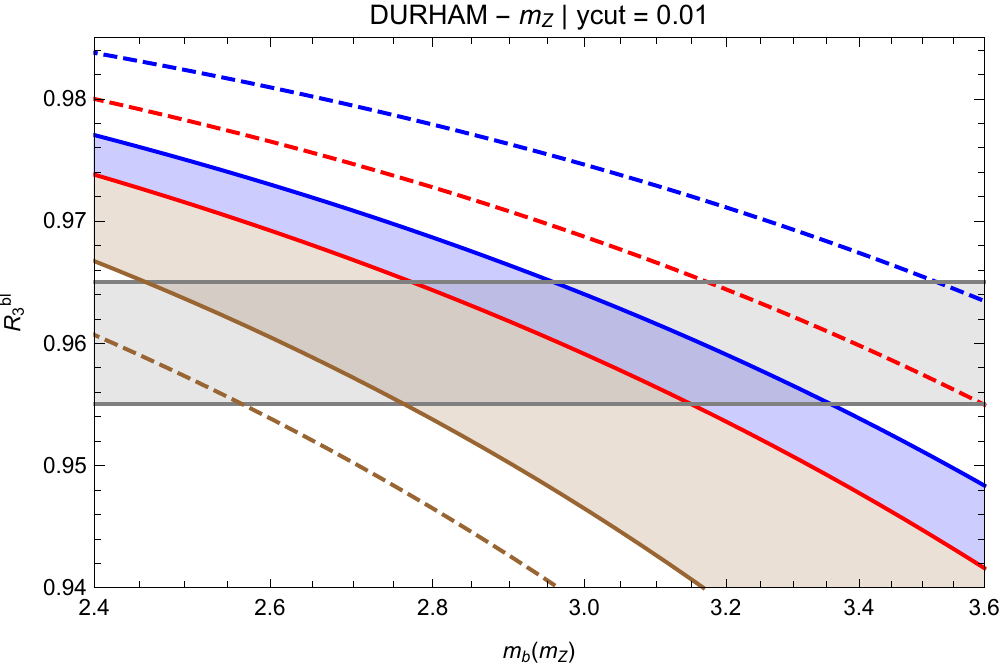}
\includegraphics[scale=.75]{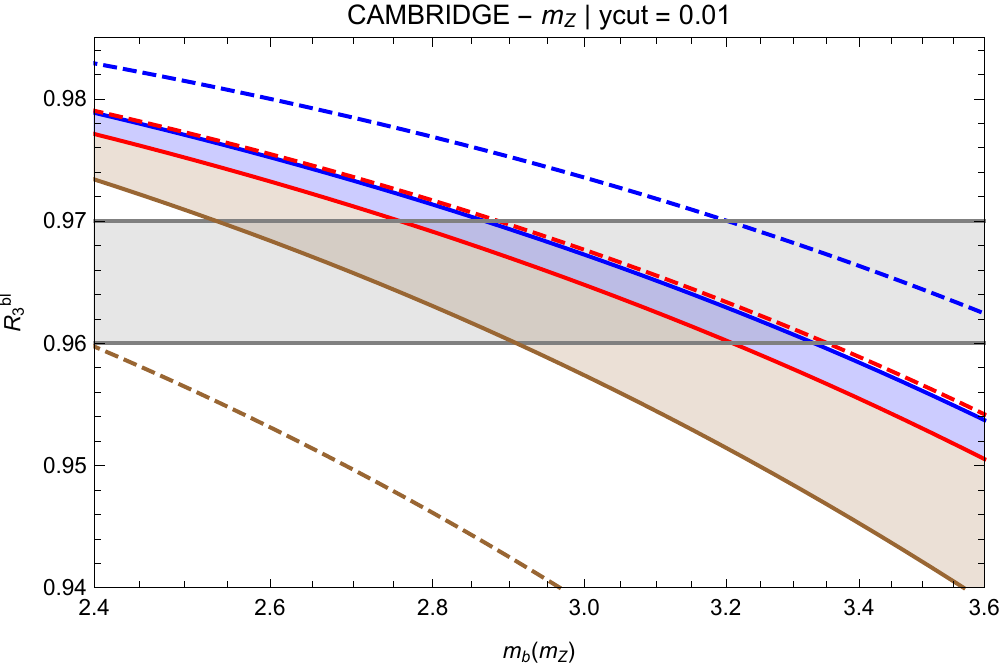}
\includegraphics[scale=.75]{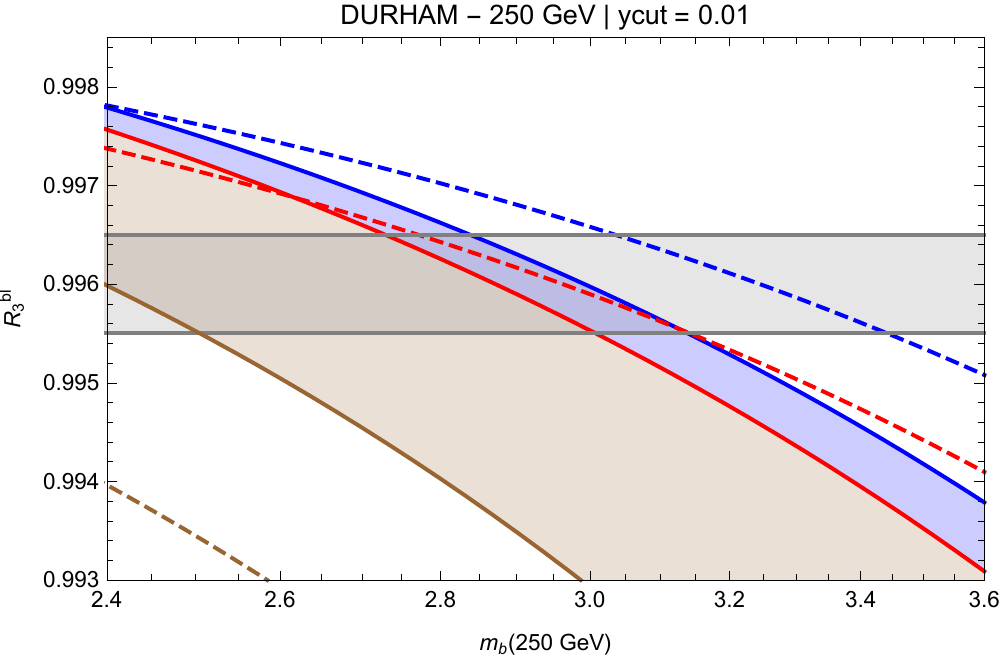}
\includegraphics[scale=.75]{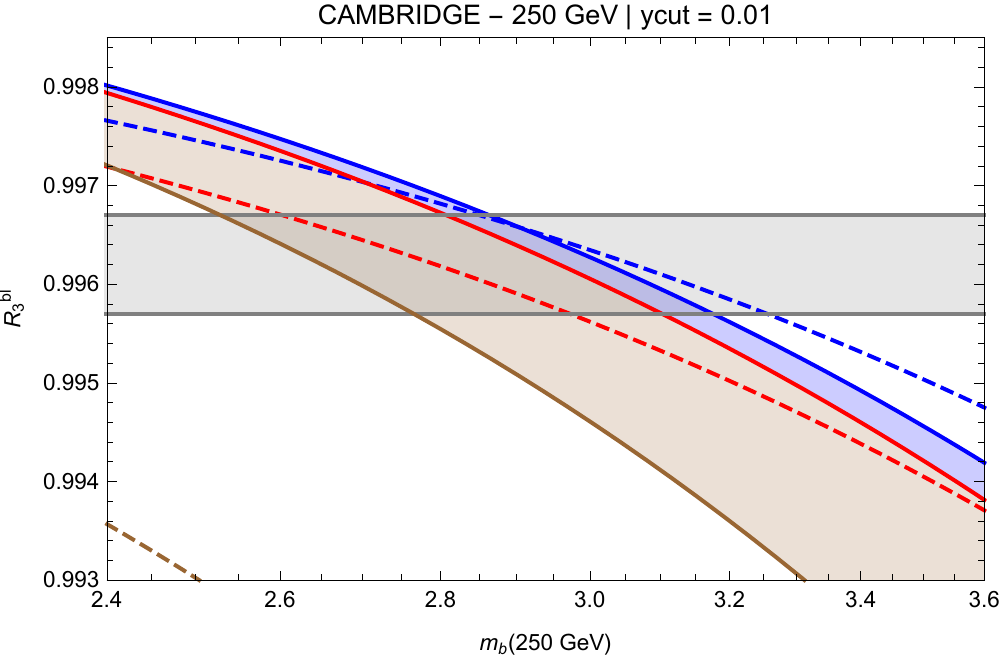}
\caption{Mass dependence of \Rbl. Same colour scheme as in  Fig.~\ref{fig:theoryr3bl}. The horizontal band represents an Ansatz for the experimental measurement.} 
\label{fig:theorymb}
\end{figure}

In Fig.~\ref{fig:theoryr3bl} we show the theory predictions for \Rbl.
The experimental uncertainty in \Rbl necessary to extract the \bquark mass value with a given uncertainty can be estimated through the approximation
\begin{equation}
    \Delta \Rbl \sim \frac{2(1-\Rbl)}{m_b(\mu)} \, \Delta m_b(\mu)~.\label{eq:sensitivity}
\end{equation}
This means, for example, that for a target uncertainty of $\Delta m_b(\mu) = 0.2~\gev$, and given the results in Fig.~\ref{fig:theoryr3bl}, we need to reach a $0.5\%$ accuracy in \Rbl at the $Z$-pole and a challenging $0.5$ per mile at $\sqrt{s} = 250~\gev$. This is consistent with Fig.~\ref{fig:theorymb} where we show the mass dependence of the theory predictions for \Rbl.

The theoretical uncertainty is estimated by considering the following sources:
\begin{itemize}
\item
Renormalization scale: The renormalization scale $\mu$ in the theoretical expression \eqref{eq:r3running} is varied from $\mu = \sqrt{s}/2$ to $\mu = 2 \sqrt{s}$ and half of the difference between the results obtained on $m_b(\mu)$ is taken as the renormalization scale uncertainty;
\item Mass ambiguity: Starting from the NLO calculation of \Rbl in terms of the pole mass $M_b$, the value of $M_b$ could be extracted and transformed to $m_b(M_b)$ which was later evolved to $m_b(m_Z)$ by means of the RGE. This is also a valid procedure to extract $m_b(m_Z)$. At infinite orders the result derived in this way and the one obtained directly from the original NLO calculation in terms of the running mass should be the same. The difference between the results obtained from the two procedures was then considered as a conservative indication of the size of the unknown higher order corrections;
\item
Strong coupling: the strong coupling $\alpha_S(m_Z)$ from \eqref{eq:input} is varied within its uncertainty. The spread of values obtained for $m_b(\mu)$ is added to the theory uncertainty. It provides a very small contribution in comparison to the renormalization scale uncertainty.
\end{itemize}

These uncertainty are evaluated quantitatively in Section~\ref{sec:prospects}.

\section{Monte Carlo simulations}

The projection for the bottom quark mass measurement is obtained from a detailed Monte Carlo simulation of the 250~\gev run at the ILC.
All results reported in this note are obtained using Monte Carlo samples
generated by the ILD concept group~\cite{Behnke:2013lya}.
This work uses the same samples, signal definitions and reconstruction tools as a previous study described in \cite{irles_poeschl_richard}.

The samples consist of events generated at leading order in QCD using the WHIZARD 1.95~\cite{Kilian:2007gr,Moretti:2001zz} event generator. The parton showering and hadronization are simulated by the Pythia 6.422 event generator~\cite{Sj_strand_2006}. QED initial state radiation (ISR) is simulated on top of the matrix element via CIRCE \cite{Ohl:1996fi}. QCD final state radiation (FSR) is implemented by Pythia in the parton shower stage.

The LO generator setup lacks the formal precision to predict the value of \Rbl accurately. The Monte Carlo simulation is primarily used to assess the importance of hadronization and experimental effects. The central values of estimated observable is re-scaled, where needed, to the Next-to-leading order (NLO) QCD prediction described in Section~\ref{sec:theory}. Studies on the implementation of NLO calculations in the WHIZARD 2 generator are ongoing, in collaboration with the Whizard authors, and may eventually be incorporated in the ILD simulation framework, but is left for a future publication.

All samples, signal and backgrounds, are generated assuming a longitudinal beam polarization of 100\%. The configuration where the electron beam is left polarized and the positron beam is right-polarized is indicated as \eLpR. The opposite configuration is labelled as \eRpL. 
The statistics of the samples is limited and corresponds to integrated luminosity of 250~\ifb 
for the signal events.

The ILC data taking program (ILC H20 program \cite{Barklow:2015tja}) foresees a total integrated 
luminosity of 2000~\ifb{} shared between different
beam polarizations schemes. Most of the luminosity will be collected
in two samples of equal size of  900 \ifb with
the polarisation schemes \mpp and \pmp.
The \mpp scheme stands for $Pol(e^{-})=P=- 80\%$ and $Pol(e^{+})=P^{\prime}=+30\%$
and the \pmp schems for $Pol(e^{-})=P=+ 80\%$ and $Pol(e^{+})=P^{\prime}=-30\%$.
This is the so called H20 luminosity scenario for the ILC.

The signal, $\ee\rightarrow Z^{*}/\gamma \rightarrow\qqbar$, cross section for different scenarios with fully 
polarised beams are listed in Table \ref{tab:crosssection}.
The cross sections of the processes contributing to the background 
contamination are listed in Table \ref{tab:crosssection_bkg}, also for 
fully polarised beam scenarios. These backgrounds come from
two different types of process. The first one is the so-called
radiative return: when the energy of the ISR, $\Kgamma$ 
is large enough to produce
a shift on the centre of mass energy of the hard interaction 
to produce a on-shell \Zboson in the process, $\ee\gamma\rightarrow Z\gamma \rightarrow \gamma\qqbar$. Therefore, the \qqbar
in the final state are not produced at 250~\gev but at the \Zpole.

\begin{table}[!ht]
  \centering
  \begin{tabular}{c|ccc}
    \hline
    Polarization & \multicolumn{3}{|c}{ $\sigma_{\eeqq} (\Kgamma<50\,\GeV)$[fb]}  \\
    \hline
    & \bbbar & \ccbar & \qqbar ($q=uds$)  \\
    \hline
    \eLpR & 5970.9 & 8935.2 & 19347.6  \\
    \eRpL & 1352.1 & 3735.1 & 5920.4 \\
    \hline
    \end{tabular}
    \caption{LO production cross section of quark pairs at 250~\GeV of centre of mass using polarized beams. \label{tab:crosssection}}
\end{table}

\begin{table}[!ht]
 \centering
  \begin{tabular}{c|c|c}
    \hline
	Channel & $\sigma_{\eLpR}$ [fb] & $\sigma_{\eRpL}$ [fb]   \\
      \hline
      $\ee\rightarrow Z\gamma \rightarrow \gamma \qqbar (\Kgamma>50\,\GeV)$ & 94895.3  &  60265.3 \\
      $\ee\rightarrow WW \rightarrow q_{1} \bar{q_{2}} q_{3} \bar{q_{4}}$ & 14874.4 & 136.4  \\
      $\ee\rightarrow ZZ \rightarrow q_{1} \bar{q_{1}}q_{2} \bar{q_{2}}$ & 1402.1 & 605.0 \\
      $\ee\rightarrow HZ \rightarrow q_{1} \bar{q_{1}}q_{2} \bar{q_{2}}$ & 346.0 & 222.0  \\
      \hline
  \end{tabular}
  \caption{\label{tab:crosssection_bkg} Cross sections at 250 \GeV for processes producing at least a pair of quarks (all flavours except $t$-quark). }
\end{table}

The ILD detector geometry and the interaction of the particles 
with the detector are simulated
within the Mokka framework interfaced with the GEANT4 toolkit 
\cite{Agostinelli:2002hh,Allison:2006ve,Allison:2016lfl}. 

\section{Event selection}
\label{sec:selection}

In this section, the main steps of the event selection are introduced.

\textbf{Low-level object reconstruction}

The simulated events are reconstructed using the standard \texttt{ILCSoft} toolkit, which includes algorithms for tracking~\cite{Gaede:2014aza} and vertex reconstruction, the Pandora particle flow algorithm~\cite{Marshall:2012hh}, photon identification, jet clustering and flavour tagging algorithms. We refer to the description in Ref. \cite{irles_poeschl_richard} for details and only describe the jet clustering and flavour tagging steps in some detail. 

The  Particle Flow Objects reconstructed with the Pandora algorithm form the input to the jet clustering algorithm. For the initial jet clustering we use the generalised $k_{t}$ algorithm for \ee colliders as implemented in \texttt{FastJet} \cite{Cacciari:2011ma}, with parameters $R=1.25$ and $p=1$, following the optimization in \cite{irles_poeschl_richard}. The algorithm is configured to reconstruct exactly two jets (exclusive clustering). 


\textbf{Flavour selection}.
The reconstruction of \Rbl requires the identification of the flavour of quarks in the $e^+e^- \rightarrow \qqbar (g)$  process.
To this end the nominal ILD flavour tagging algorithm is applied on both jets. The \texttt{LCFIPlus} package~\cite{Suehara:2015ura} provides a vertex reconstruction algorithms. The jet flavour tag is assigned using boosted decision trees (BDTs) that take into account a number of
variables from tracks and vertices. For each jet the \texttt{LCFIPlus} algorithm returns two tagging variables, $btag$ and $ctag$, which reflect the likelihood that the jet originated
from the fragmentation of a \bquark or a \cquark, respectively. The likelihood distribution for the $e^+e^- \rightarrow \qqbar (g)$ at $\sqrt{s}=$ 250~\gev{} is shown in Fig.~\ref{fig:flavour_tagging}, for the \eLpR configuration in the left panel and for the \eRpL configuration in the right panel.
The contributions from the different processes, classified by the Monte Carlo truth information, are indicated with different fill colours in the histograms.

Events are assigned to the \bquark or \lquark sample based on a double tag, requiring that both jets satisfy the following criterion:  
\begin{itemize}[leftmargin=1.0in]
        \item For the \bquarks sample: $btag>0.85$
        \item For the \lquark sample: $btag<0.4$ \&  $ctag<0.25$
\end{itemize}

\begin{figure}[!pt]
  \centering
      \begin{tabular}{cc}
        \includegraphics[width=0.45\textwidth]{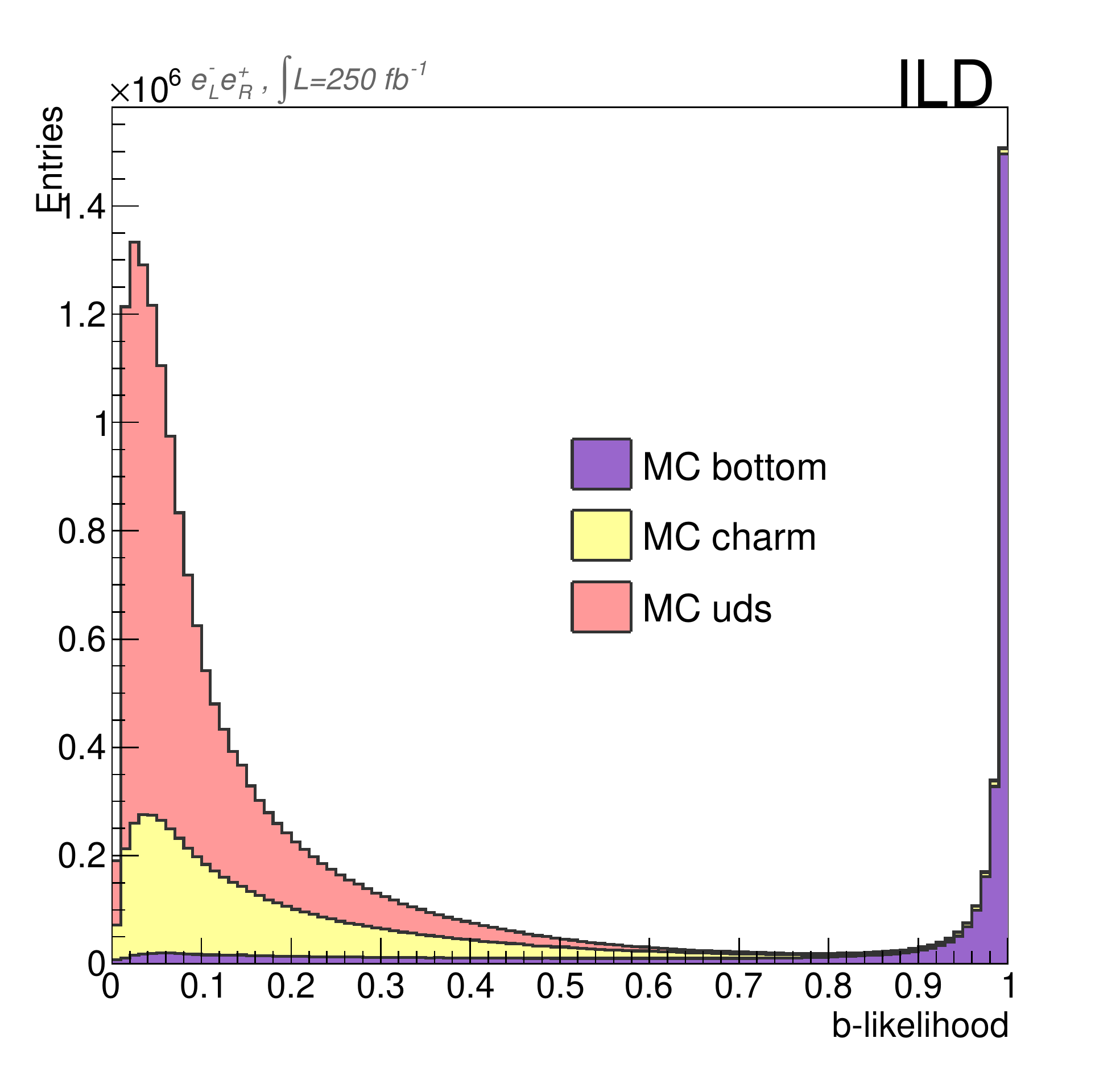} &
        \includegraphics[width=0.45\textwidth]{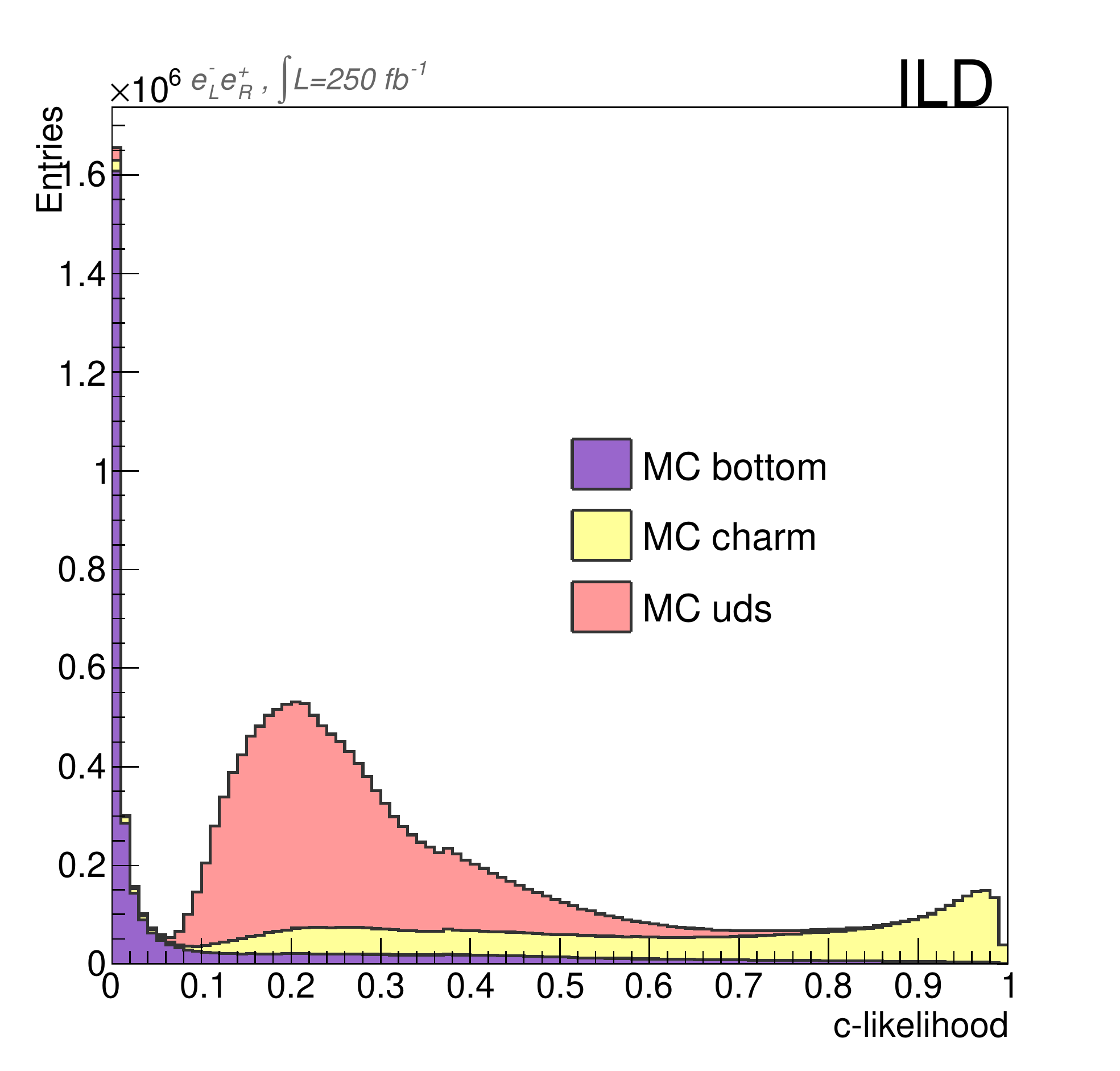}
      \end{tabular}
      \caption{\label{fig:flavour_tagging}Distributions of b-likelihood(left) and c-likelihood(right) for left polarised \(e^+e^-\to q\overline{q}\) sample.}
\end{figure}

The efficiency and purity of the flavour tag classification are summarized in Table~\ref{tab:tagging} for \bquarks and \lquarks. For comparison, the performance of the DELPHI selection is shown in the same table. The efficiency with which ILD expects to select the \bquark sample is approximately 80\%, with a purity of nearly 98\%. It thus exceeds the DELPHI results in both aspects very considerably, as might be expected from the advance in vertex detector technology since LEP era together with 
a more precise primary vertex reconstruction allowed by the higher energy of the jets and the excellent vertexing due to the smaller radius of
the vacuum pipe. Also for the \lquark sample the $b$ and $c$-quark vetoes are expected to work somewhat more efficiently, and with much greater purity, than in the DELPHI analysis.

\begin{table}[!ht]
 \centering
  \begin{tabular}{c|cc|cc}
    \hline
	  & \multicolumn{2}{c|}{\bquarks} & \multicolumn{2}{c}{\lquarks} \\
	 Experiment & Eff. [\%] & Pur. [\%] & Eff. [\%] & Pur. [\%] \\
      \hline
    DELPHI \cite{Abdallah:2005cv} & 47\% & 86\% & 51\% & 82\% \\
    ILD (this note) & 80\% & 98.7\% & 58\% & 96.1\% \\
      \hline
  \end{tabular}
  \caption{\label{tab:tagging} Prospects for efficiency and purity of single jet tagging at ILD compared with the values published in \cite{Abdallah:2005cv} used as reference of the \Rbl experimental method. }
\end{table}

\textbf{Rejection of radiative return events}. To clearly identify the scale of the process used to measure $m_b(m_Z)$ events where a hard photon from Initial State Radiation (ISR) reduces the center-of-mass energy significantly are removed from the sample. The simplest way to remove these radiative-return events is to veto events with an energetic ISR photon. We therefore veto events with an energetic photon. However, for a large fraction of the cases, ISR photons are emitted at low angle and often the photon is lost
in the beam pipe. Again following Ref.~\cite{irles_poeschl_richard},
we reject these events using angular and energy conservation criteria. For this analysis, cuts are applied on the invariant mass of the two-jet system and on the \Kreco quantity. This quantity
is used as an estimator of the momentum of the ISR, $|\vec{k}|$ and it is defined as:
\begin{equation}
    |\vec{k}| \approx \Kreco  = \frac{250\, \text{GeV} \cdot \sin{\Psi_{acol}}}{\sin{\Psi_{acol}}+\sin{\theta_{1}}+\sin{\theta_{2}}}
\end{equation}
where the $\Psi_{acol}$ variable describes the acolinearity between the two reconstructed jets:
\begin{equation}
    \sin{\Psi_{acol}}=\frac{|\vec{p}_{j_{1}}\times \vec{p}_{j_{2}}|}{|\vec{p}_{j_{1}}|\cdot|\vec{p}_{j_{1}}|}
\end{equation}
and $\theta_{j}$ refers to the reconstructed polar angle of the jet-$j$ in the detector reference frame.

The cuts that aim to reduce the contamination by radiative-return events can be summarized as follows:
\begin{itemize}[leftmargin=1.0in]
    \item no photons with $E_\gamma >$ 100~\gev \item no photon with $E_\gamma >$ 70~\gev{} and $|\costheta|>$ 0.95
    \item $\Kreco<50$ GeV 
    \item $m_{2jets}>130$ GeV.
\end{itemize}

This selection removes approximately 99\% of the radiative-return events, and a good fraction of the diboson background events, while keeping around 80\% of our signal events.

\textbf{Rejection of the boson pair production background}. After the radiative-return and flavour selection, a significant background due to boson pair production still remains. In Ref.~\cite{irles_poeschl_richard}, this background \cite{irles_poeschl_richard} is rejected with a cut on the $y_{23}$ variable which is tightly connected to three-jet rates.
To avoid a strong bias in the \Rbl measurement, we investigate other
event shapes variables, motivated by the selection procedure described
in \cite{Abdallah:2005cv}.
In Fig.~\ref{fig:preselection_thrust} we show the potential of the $Thrust$
variable to remove background events. This variable is defined as:
\begin{equation}
Thrust= \max\limits_{|\vec{n}|=1} (\frac{\sum\limits_{PFO}|\vec{n}\cdot\vec{p}_{PFO}|}{\sum\limits_{PFO} |\vec{p}_{PFO}|})
\end{equation}

A mild cut is found to reduce the boson pair background strongly, with a minimal bias on the signal events.
We therefore apply the following $WW/ZZ/ZH$ rejection cut:
\begin{itemize}[leftmargin=1.0in]
        \item $Thrust$ > 0.8.
\end{itemize}

\begin{figure}[!h]
  \centering
      \begin{tabular}{cc}
        \includegraphics[width=0.45\textwidth]{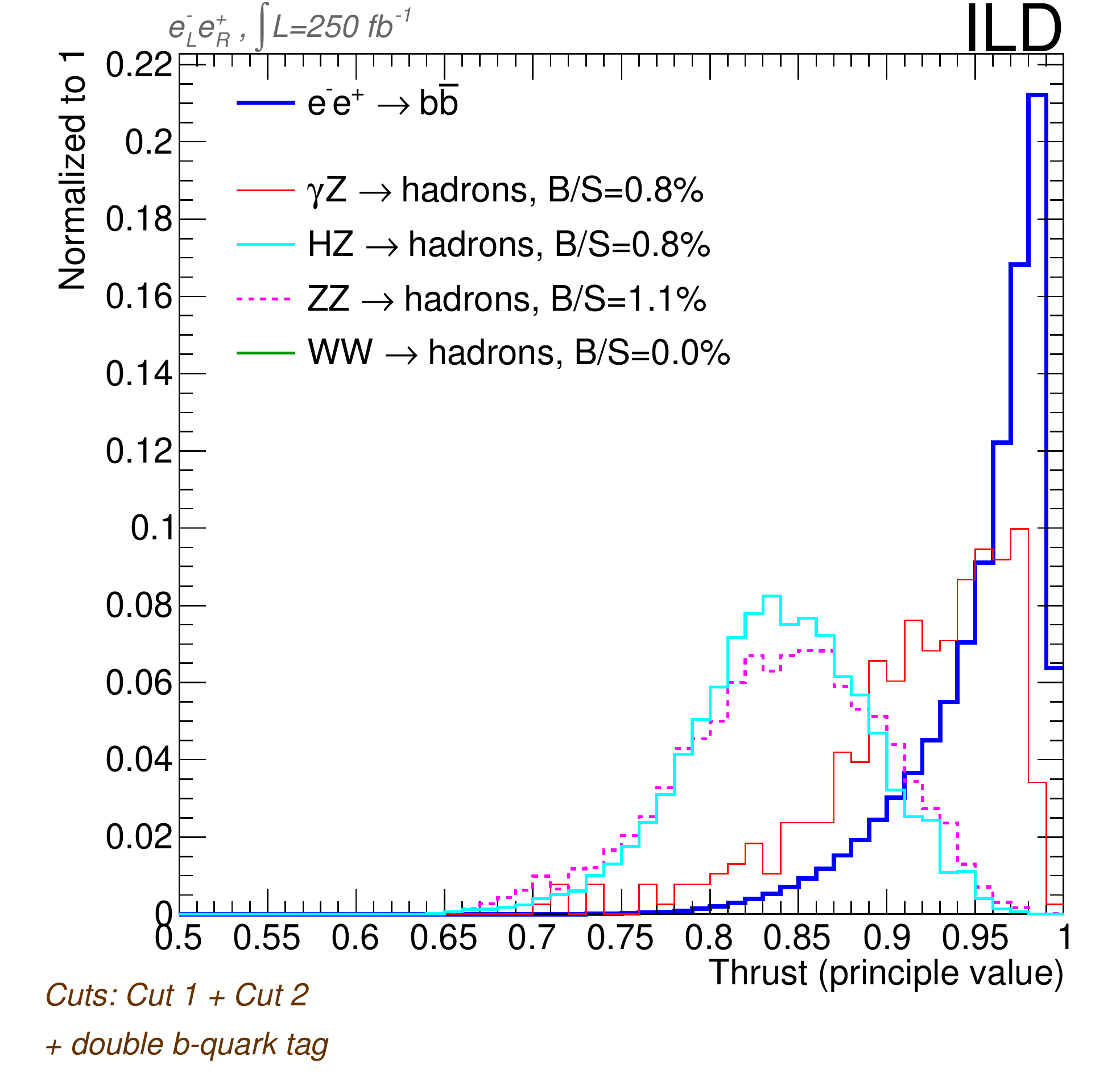} &
        \includegraphics[width=0.45\textwidth]{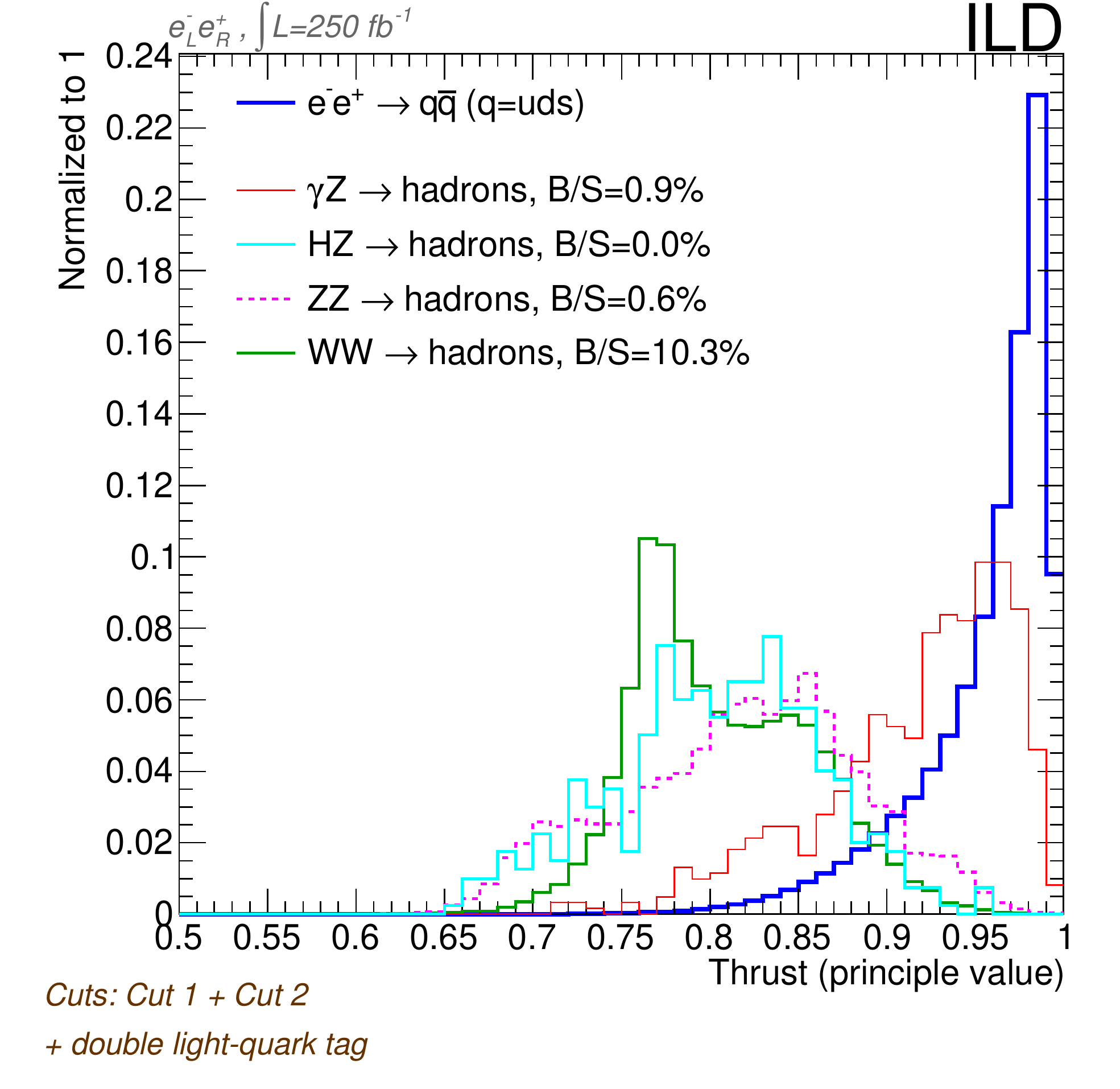} \\
        \includegraphics[width=0.45\textwidth]{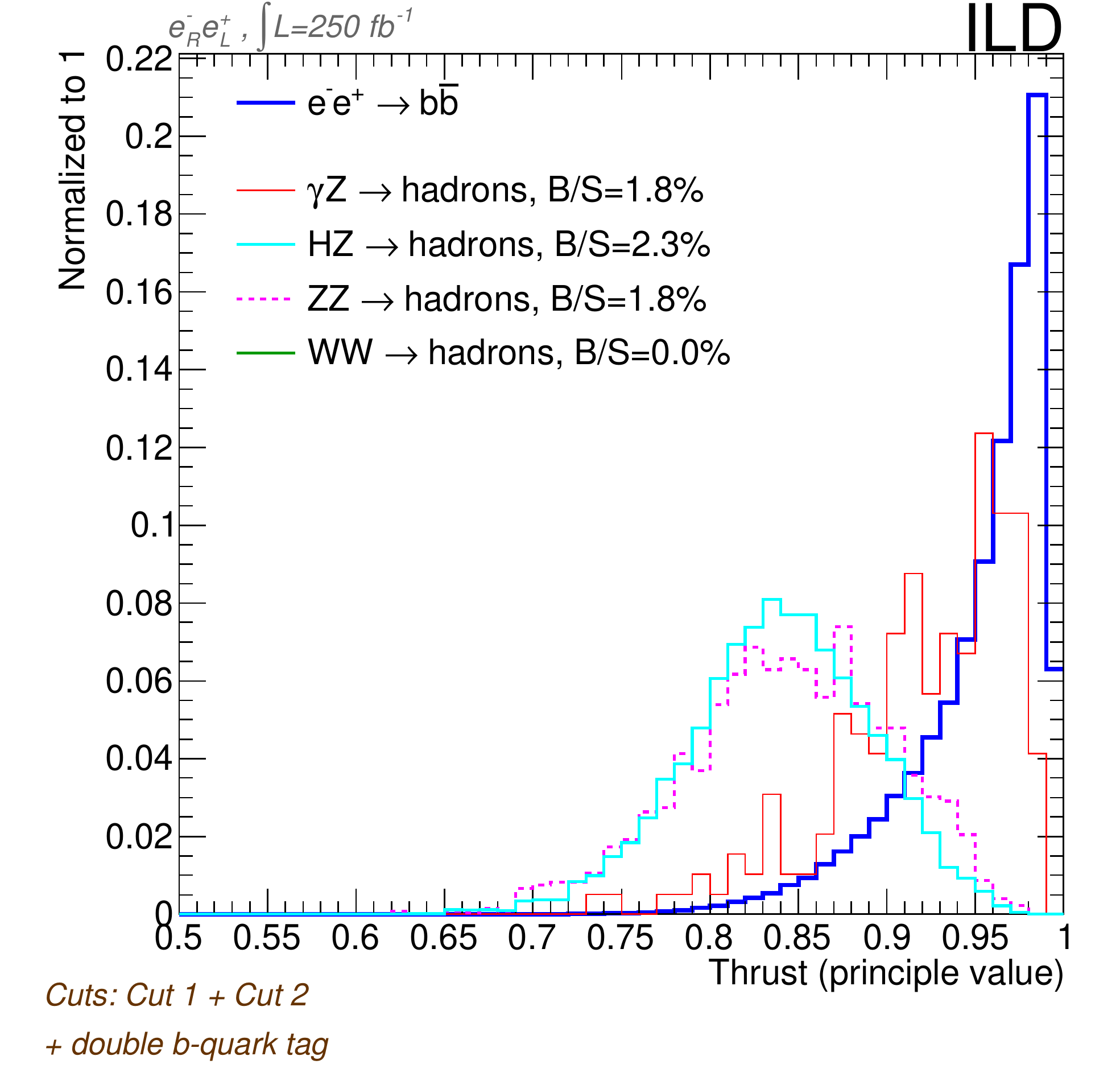} &
        \includegraphics[width=0.45\textwidth]{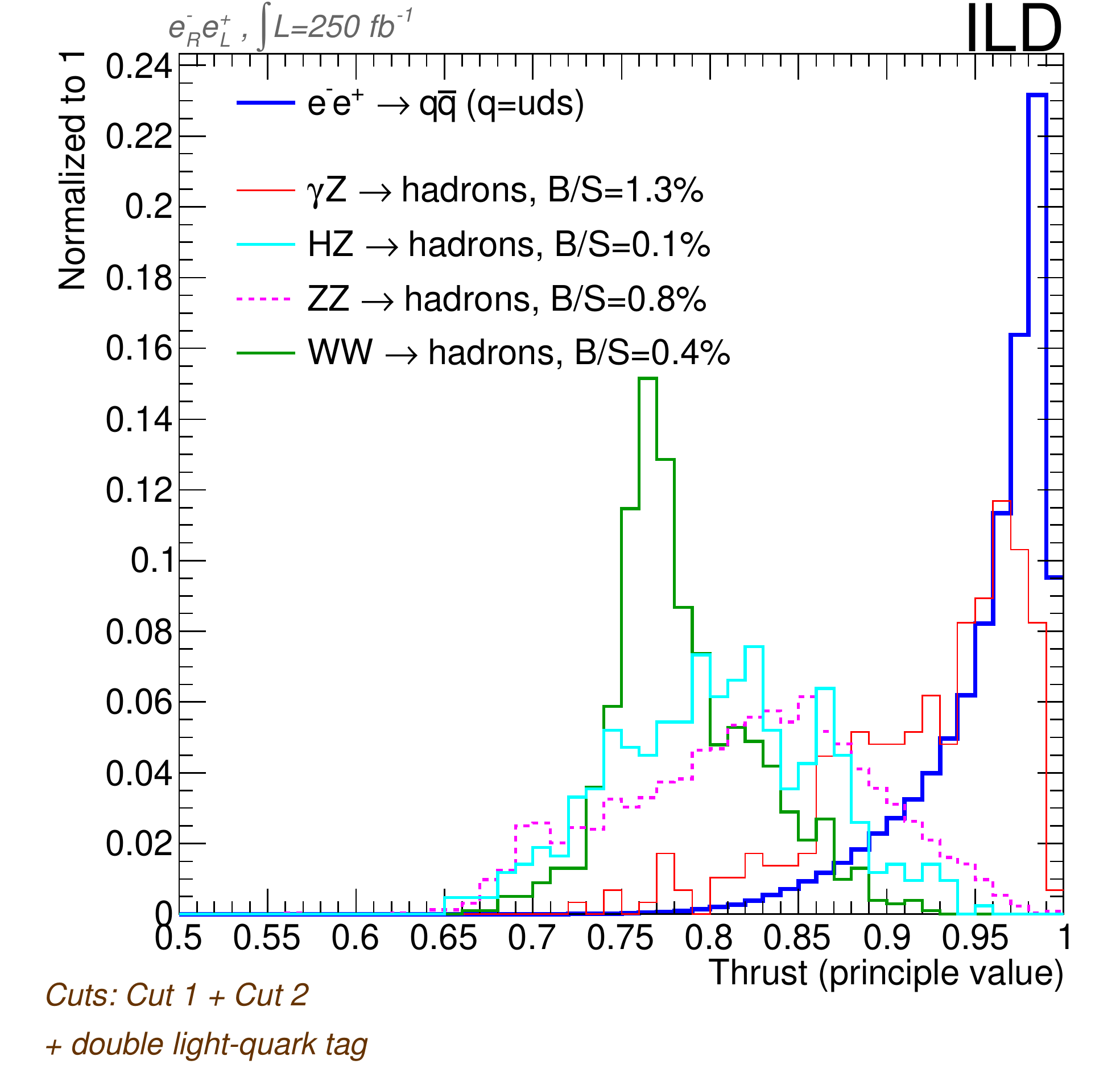} \\
      \end{tabular}
      \caption{\label{fig:preselection_thrust} Distribution of the Thrust variable used to discriminate the remaining di-boson backgrounds from the signal events. The upper row presents the results for the \eLpR configuration, the lower row for \eRpL. The plots in the leftmost panel are for \bquark production, the ones in the rightmost panel for \lquark (uds) production.}
\end{figure}

\textbf{Selection efficiency and purity}. The result of the selection procedure described above is summarized in Table~\ref{tab:cutflow}. The efficiency for the $e^+e^- \rightarrow b\bar{b}(g)$ process is nearly 40\% for both beam polarizations. The \lquark sample is selected with an efficiency slightly over 16\%. The contamination by radiative-return events, that complicate the assignment of an unambiguous scale to the process, is controlled to the 1-2\% level in both samples. The contribution to the event yield of other processes is expected to be order of 2-5\% for the \lquark sample in the \eLpR configuration, and much smaller for the \bquark sample and the \eRpL beam polarization configuration.

\begin{table}[!ht]
 \begin{center}
  \begin{tabular}{c|c|cccc}
    \hline
    \hline
     \multicolumn{6}{c}{ {\textbf \eLpR} }\\
    \hline
    \hline
      & & \multicolumn{4}{c}{ B/S }\\
      \hline
      & {\textbf Signal Eff [\%]} & Rad. Return& $WW$ & $ZZ$ & $HZ$ \\
      \hline
          \multicolumn{6}{l}{T>0.8} \\
    \hline
      \Rl & 16.5\% & 1.4\% & 5.1\% & 0.3\% & 0.0\%\\
      \Rb & 37.8\% & 1.2\% & 0.0\% & 0.6\% & 0.6\%\\
   \hline
             \multicolumn{6}{l}{T>0.85} \\
    \hline
      \Rl & 16.2\% & 1.3\% & 2.3\% & 0.2\% & 0.0\%\\
      \Rb & 36.9\% & 1.2\% & 0.0\% & 0.3\% & 0.3\%\\
   \hline
    \hline
    \\
    \\
        \hline
    \hline
     \multicolumn{6}{c}{ {\textbf \eRpL} }\\
    \hline
    \hline
      & & \multicolumn{4}{c}{ B/S}\\
      \hline
      & {\textbf Signal Eff [\%]} & Rad. Return& $WW$ & $ZZ$ & $HZ$ \\
      \hline
          \multicolumn{6}{l}{T>0.8} \\
    \hline
      \Rl & 16.7\% & 1.5\% & 0.1\% & 0.5\% & 0.0\%\\
      \Rb & 37.3\% & 1.9\% & 0.0\% & 1.4\% & 1.8\%\\
   \hline
             \multicolumn{6}{l}{T>0.85} \\
    \hline
      \Rl & 16.4\% & 1.4\% & 0.0\% & 0.3\% & 0.0\%\\
      \Rb & 36.5\% & 1.8\% & 0.0\% & 0.9\% & 1.0\%\\
   \hline
    \hline
  \end{tabular}
  \caption{\label{tab:cutflow} Cut flow for the signal and background events.}
 \end{center}
\end{table}

\section{Measurement of \Rbl at $\sqrt{s}=$ 250~\gev}
\label{sec:r3bl}

\textbf{Definition of the observable}. The \Rbl observable is reconstructed by re-clustering the Particle Flow Objects in the selected events using the Cambridge algorithm~\cite{Dokshitzer:1997in}. This algorithm was used in the most precise DELPHI measurement, where it was shown to reduce the hadronization uncertainties compared to the Durham algorithm. In principle, one could explore the use of more robust algorithms~\cite{Boronat:2016tgd}, but as the NLO calculation of Section~\ref{sec:theory} is available for Cambridge, we stick to the Cambridge algorithm in the current study.

For a given flavour, $q$, after the full selection procedure we get a sample of events forced to be clustered as two jets and with both of them tagged as originated from a $q$-quark.
The size of this sample is $N^{q}_{double-tag}$.
Studying the internal structure of the jets using the Cambridge algorithm we determine how many of these events would be reconstructed as 3 jets, as a function of $y_{cut}$. The size of this sample is $N^{q,3-jets}_{double-tag}$. With these two distributions we define the three-jet rate as follows:
\begin{equation}
    \left. \Rq \right|_{reco} = \frac{N_{double-tag}^{q,3-jets}(y_{cut})}{N_{double-tag}^{q}}.
\end{equation}
where $q$ indicates the quark flavour, and can take two values: $q=b$ or $q=\ell$. The double ratio $\left. \Rbl \right|_{reco}$ is the ratio of $\left. \Rb \right|_{reco}$ and $\left. \Rl \right|_{reco}$, as in Eq.~\ref{eq:r3bl}.


\textbf{Monte Carlo results for \Rbl}. The value of \Rbl in signal events is shown as a function of $y_{cut}$ in Fig.~\ref{fig:R3bl}. 

\begin{figure}[!h]
  \centering
     \includegraphics[width=0.6\textwidth]{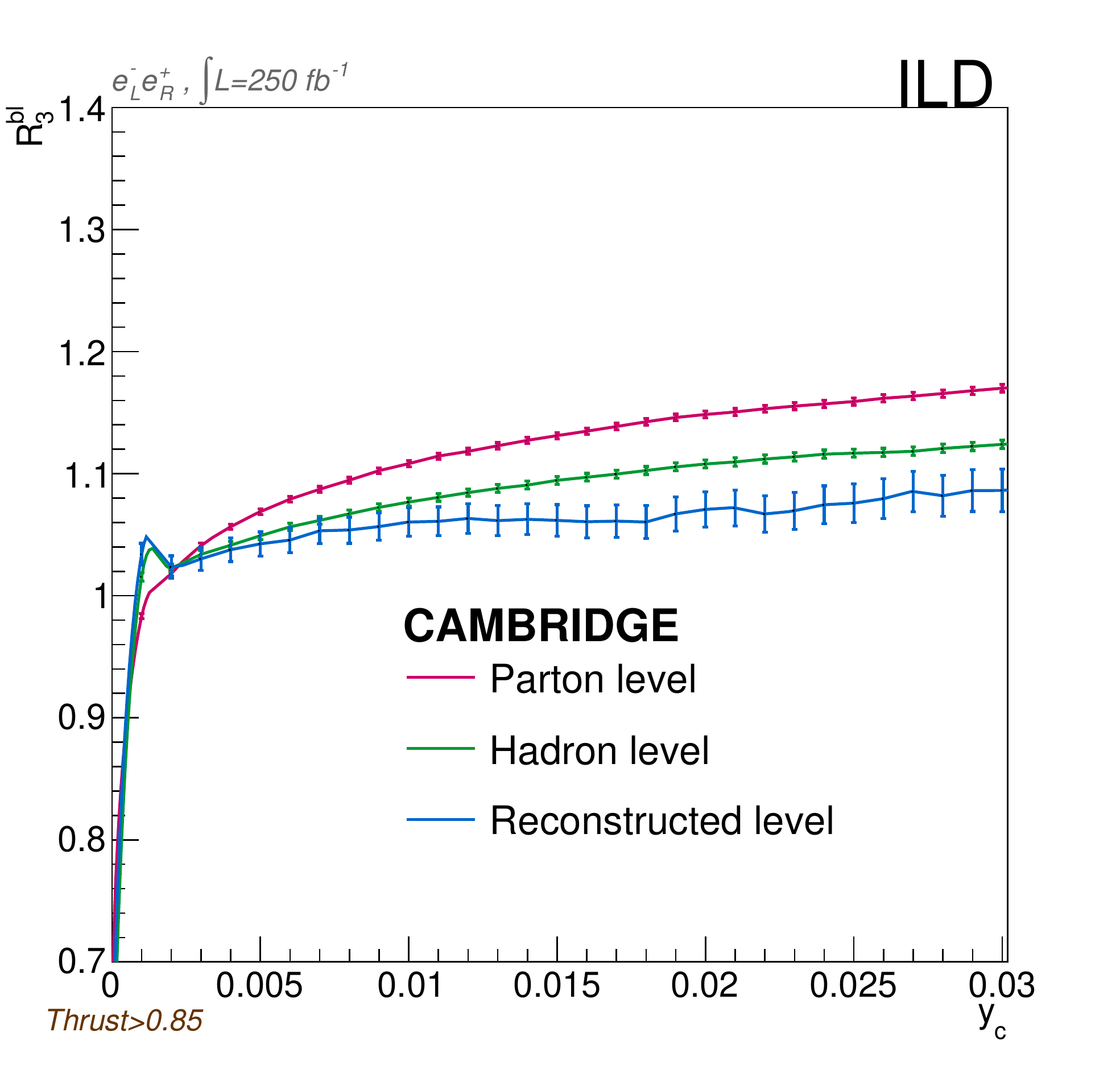} 
      \caption{\label{fig:R3bl} \Rbl distribution at parton level (red), at hadron level (green) and at the level of reconstructed objects (dark blue) applying the selection cuts described in the previous section. All curves correspond to signal events, without backgrounds. The central value for \Rbl predicted by the LO Monte Carlo is not reliable since the gluon radiation and the quark mass effects are only generated by the parton shower algorithm; only the difference between the curves are meaningful, indicating a size of the correction between generator levels. \label{fig:r3blexp}}
\end{figure}

The observable is defined at several different levels as shown in Fig.~\ref{fig:r3blexp}. The red markers, labelled as parton shower level, indicate the result that is obtained when the Cambridge clustering is applied to partons from the Pythia truth record after Final State Radiation. This level is closest~\footnote{We reiterate here that the formal accuracy of the LO Monte Carlo generator with massless b-quarks in the matrix element is insufficient to provide a reliable prediction of the true value of the observable. Only the differences between the different levels is considered meaningful in Fig.~\ref{fig:R3bl}.} to the result of the fixed-order calculation reported in Section~\ref{sec:theory}. The green markers, labelled as hadron level, indicate the result obtained when clustering stable final-state particles. The sets of points with blue markers indicate the result at the detector level, obtained when clustering particle flow objects. 

Clearly, the observable is transformed in important ways by the parton shower and the hadronization into colour-neutral particles and by the response of the detector.

\textbf{Correction to the parton level}. To compare the measured value of \Rbl to a fixed-order calculation, the result must be corrected back to parton level.
Following the same approach as Ref.~\cite{Abdallah:2005cv}, the result is unfolded with two correction factors:
\begin{equation}
    \left. \Rbl\right|_{parton} = C_{had} \times C_{det} \times \left. \Rbl \right|_{reco}
\end{equation}

The first correction factor, $C_{had}$, contains the correction to the parton level distribution, undoing the effect of hadronization modelling. 

The second factor, $C_{det}$, corrects the effect of the detector resolution and any bias introduced by the detector acceptance and efficiency. It can be broken down in terms of the signal selection efficiency $\epsilon_{sel}$, the background selection efficiency $\epsilon_{bkg}$ and the flavour tagging efficiency $\epsilon_q$ and mis-tag probability $\epsilon_q'$:
\begin{equation}
\left. \Rq(y_{cut}) \right|_{reco}=
  \frac{\epsilon_{sel} \cdot \left[ \epsilon_{q}^{2} \sigma_{\qqbar}^{3jet}(y_{cut}) +  \epsilon_{q'}^{2} \sigma_{q'\bar{q}}^{3jet}(y_{cut}) \right]  + \epsilon_{bkg} \sigma_{bkg}^{3jet}(y_{cut})} {\epsilon_{sel} \cdot \left[ \epsilon_{q}^{2} \sigma_{\qqbar} +  \epsilon_{q'}^{2} \sigma_{q'\bar{q}} \right]  + \epsilon_{bkg} \sigma_{bkg}}.
  \label{eq:propagationsyst}
\end{equation}

The values obtained from the Monte Carlo simulation for these correction factors are shown in the Fig.~\ref{fig:correction}, separately for the two polarization configurations and for the signal and background samples.

\begin{figure}[!pt]
  \centering
      \begin{tabular}{cc}
        \includegraphics[width=0.45\textwidth]{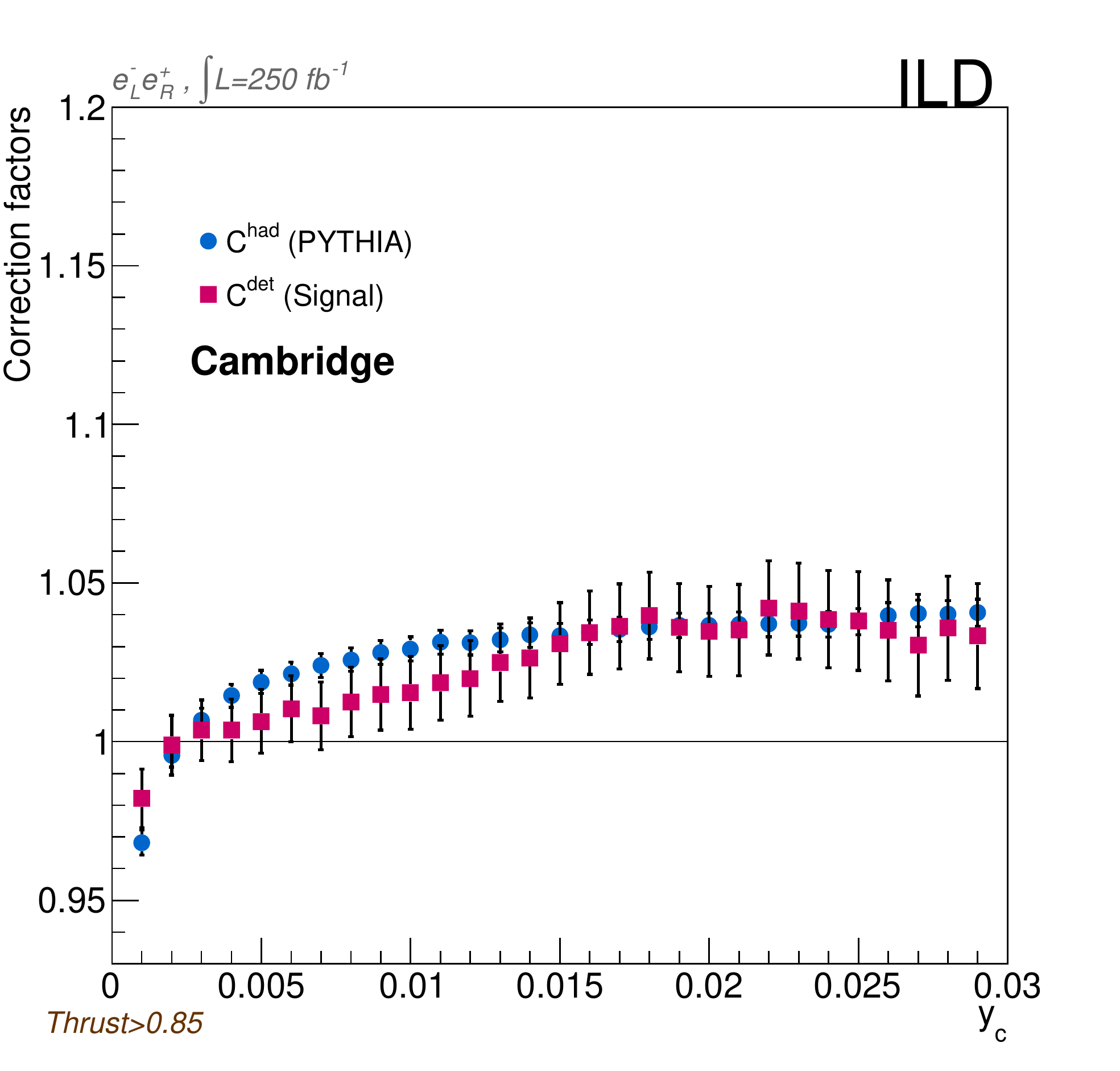} 
        \includegraphics[width=0.45\textwidth]{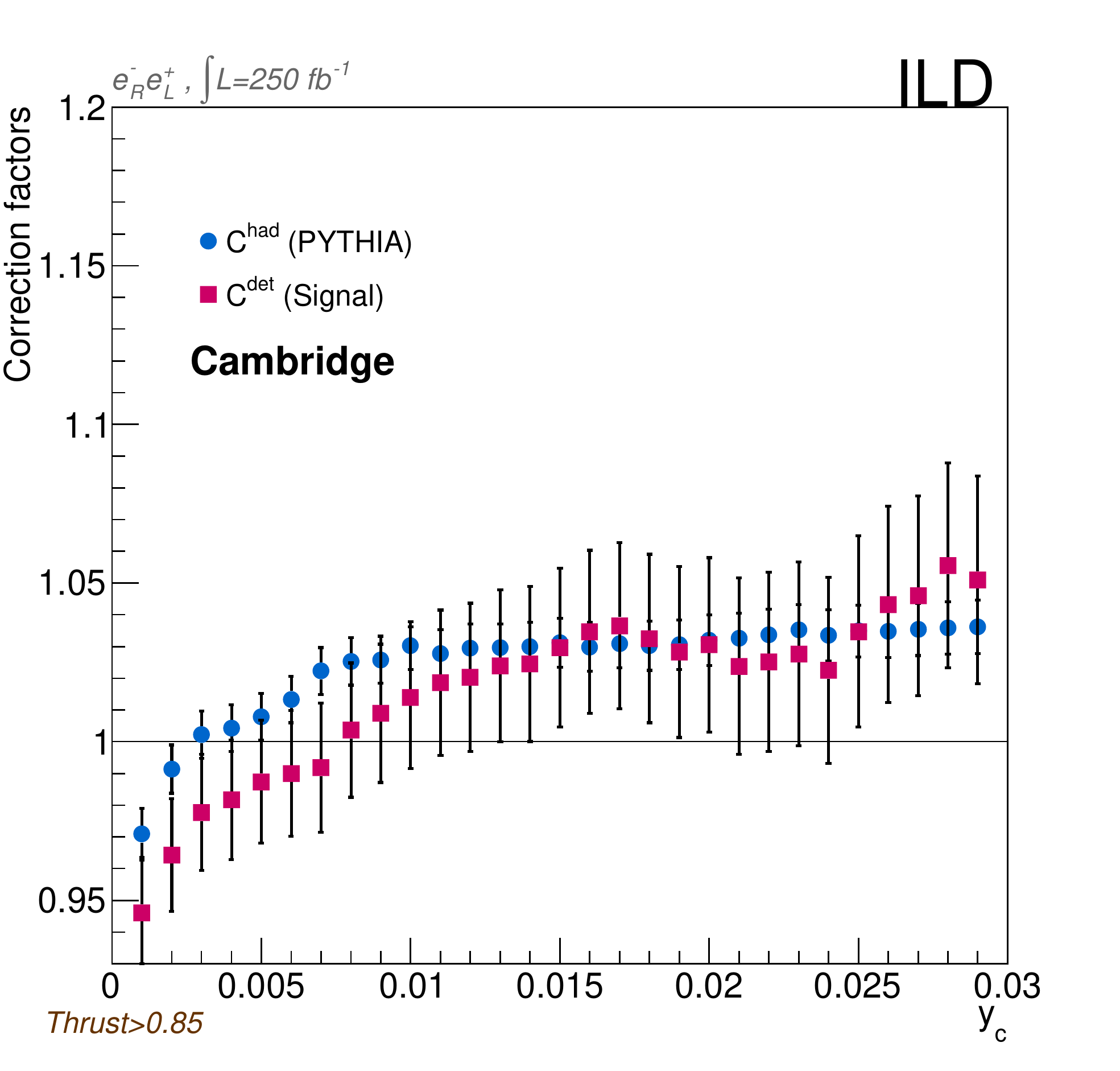} 
      \end{tabular}
      \caption{\label{fig:correction} \(C^{had}\) and \(C^{det}\) for only signal and the two $100\%$ beam polarisation cases used in our simulated samples.}
\end{figure}


For a choice of $y_{cut}=$ 0.01, the correction for hadronization and detector effects are of the order of 2-3\%. The correction for the background samples is found to be very similar to that of the signal. The systematic uncertainty on the correction is expected to be a fraction of the correction. We estimate these systematic uncertainties in the following.

\textbf{Statistical uncertainty}. The statistical uncertainty of the \Rbl measurement with an integrated luminosity of 2~\iab{} at $\sqrt{s}=$ 250~\gev{} is, combining both channels, of 1~\gev{}, as explained in Section~\ref{sec:prospects}.

\textbf{Estimation of systematic uncertainties}. The main sources of uncertainty in the DELPHI analysis of Ref.~\cite{Abdallah:2005cv} correspond to the hadronization
modelling and the contamination of other quarks in the double tagged samples. We proceed to estimate the dominant systematic uncertainties for the measurement at 250~\gev{}.

\textbf{Flavour tagging}. As shown in the Table \ref{tab:tagging}, the tagging capabilities of ILD are 
expected to be far superior than those of the LEP experiments, therefore reducing the 
miss-tag rates considerably. The flavour tagging efficiencies can be determined \textit{in situ} using the double and single tag rates.  Ref.~\cite{irles_poeschl_richard} predicts a precision at
the 0.1-0.5\% level. Propagating these uncertainties to the \Rbl measurement from Eq.~\ref{eq:propagationsyst} one obtains an uncertainty of 0.07\% for the \eLpR configuration and 0.06\% for the \eRpL beam polarization.
The other stages of the event selection may also bias the result, but its effect is expected to cancel to a good extent in the double tag rate. 

\textbf{Hadronization}. The uncertainties related to the  hadronization correction are expected to be smaller than at LEP, thanks to progress in Monte Carlo generators and the important data set collected at the Tevatron and LHC since LEP era.
On general grounds, one expects that the impact of hadronization is further reduced at higher centre-of-mass energies. We assume that the this uncertainty is a half of that in the DELPHI measurement of Ref.~\cite{Abdallah:2005cv}, which leads to an 0.1\% uncertainty on \Rbl.

\textbf{Background modelling}.
In contrast with the analysis running at the \Zpole, the analysis at $\sqrt{s}=$ 250~\gev{} is sensitive to the modelling of the backgrounds.
Assuming that the experiment can control the normalization of the remaining di-boson background at the per cent level, the uncertainty on \Rbl is approximately 0.22\% for the \eLpR configuration and 0.1\% for the \eRpL measurement.

The estimates for the dominant uncertainties are summarized in Table~\ref{tab:syst}.
%


\begin{table}[!ht]
\begin{center}
\begin{tabular}{ C{3cm} | c | L{6cm} }
    \hline
    \hline
     \multicolumn{3}{c}{ $C_{had}$ Systematic Unc. }\\
         \hline
    \textbf{Source} & \textbf{ Estimation} & comments \\
    \hline
    hadronization modelling & 0.1 \%  & Assumed to be half the uncertainty evaluated for LEP \\
    \hline
    \hline
    \multicolumn{3}{c}{ $C_{det}$ Systematic Unc. (\eLpR) }\\
    \hline
    flavour tagging & 0.07 \%  &  assuming flavour tagging uncertainties as estimated in \cite{irles_poeschl_richard} \\
    pre-selection efficiency & 0.06 \%  & as estimated in \cite{irles_poeschl_richard} \\
    $Z\gamma/WW/HZ/ZZ$ modelling & 0.20 \%  & assuming modelling uncertainties at the per cent level.
    It assumes a moderated cut in the thrust of the event which may required further studies to reject possible biases on the observable due to this cut. \\
    \hline
    \textbf{total} & \textbf{ 0.22 \%} & dominated by the $WW$ contamination to \Rl \\
   \hline
   \hline
    \multicolumn{3}{c}{ $C_{det}$ Systematic Unc. (\eRpL) }\\
    \hline
    flavour tagging & 0.06 \%  & assuming flavour tagging uncertainties as estimated in \cite{irles_poeschl_richard} \\
    pre-selection efficiency & 0.06 \%  & as estimated in \cite{irles_poeschl_richard} \\
    $Z\gamma/WW/HZ/ZZ$ modelling & 0.1 \%  & Assuming modelling uncertainties at the per cent level.
    No specific cuts are needed for the removal of the backgrounds. \\
    \hline
    \textbf{ total} & \textbf{ 0.13 \%}  & dominated by the $ZZ$ and radiative return contamination to \Rb \\
    \hline
    \hline
     \end{tabular}
  \caption{\label{tab:syst} Comprehensive assessment of the expected main systematic uncertainties on the measurement of \Rbl.}
\end{center}
\end{table}
     
\section{Prospects for bottom-quark mass measurements at the ILC}
\label{sec:prospects}

In this section we present the estimates of the uncertainty on the bottom mass that can be achieved in the envisaged runs at the ILC. Apart from the main result of this note, the prospect for the measurement of $m_b(250\gev)$, we extrapolate the precision achieved at LEP to the GigaZ run.

\textbf{Measurement of $m_b(250~\gev)$ in the "Higgs factory" run}.
We estimate the precision of the bottom quark mass measurement at $\sqrt{s}=$ 250~\gev from the projection for \Rbl in Section~\ref{sec:r3bl}. 
The statistical and systematic uncertainties are estimated using the mass sensitivity of the \Rbl observable as approximated in Eq. \ref{eq:sensitivity}. A total integrated luminosity of 2~\iab{} is envisaged in the standard ILC operating scenario, with 40\% of the data collected in the LR polarization,
and 40\% in the RL configuration. The central value for \Rbl is taken to be the SM prediction, $\Rbl =$ 0.996, obtained with the NLO QCD calculation. The statistical uncertainty is estimated for a value of $ \Rb \sim \Rl \sim 0.3$ for $y^{CAMB}_{cut}=0.01$.
The systematic uncertainties on the \Rbl measurement  are propagated to the mass determination
using the information from the Eq. \ref{eq:sensitivity} and the Fig.~\ref{fig:theoryr3bl} (bottom right).
The theoretical uncertainty is estimated from variations in the renormalization scale and mass scheme and the value of the strong coupling constant in Section \ref{sec:theory}. The result is divided by two to anticipate a future NNLO calculation of the observable.



This procedure yields the following expected precision for the mass measurements performed with the two main beam polarization 
configurations:
\begin{align}
    \Delta m_{b} (\mpp) = \pm 0.85 (stat.) \pm 0.34 (had.) \pm 0.75 (exp.) \pm 0.07 (th.)~\gev \\
    \Delta m_{b} (\pmp) = \pm 1.53 (stat.) \pm 0.34 (had.) \pm 0.44 (exp.) \pm 0.07 (th.)~\gev \
\end{align}
The two results are combined with the best linear unbiased estimator procedure (BLUE~\cite{Nisius:2020jmf}), that takes into account the correlations among systematic uncertainties (100\% for hadronization and theory uncertainties, 50\% for the experimental systematics). The projected uncertainty for the mass measurement is given by:
\begin{align}    
    \Delta m_{b} (250~\gev) = 1.0~\gev = \pm 0.76 (stat) \pm 0.59 (syst.) \pm 0.34 (had.) \pm 0.07 (th.)~\gev 
\end{align} 

The precision on the high-scale is not directly competitive with measurements at lower scales, but extends the analysis to scales not probed by previous experiments.


\textbf{Measurement of $m_{b}(m_Z)$ in the "GigaZ" scenario}. The "GigaZ" run is an option in the ILC program, where an integrated luminosity of $\sim$ 100~\ifb is collected at the $Z$-pole~\cite{Barklow:2015tja}. This run is thus expected to generate two orders of magnitude more \bbbar pairs than LEP1. With the larger data set and the excellent flavour tagging performance the ILC experiments can improve the measurement of $m_b(m_Z)$ very considerably.

The statistical uncertainty is expected to become sub-dominant.
We assume that the theory uncertainties are reduced by a factor two with respect to the LEP result, anticipating an NNLO calculation of the three-jet rate. Also the hadronization uncertainty is divided by two, which requires an important improvement of the hadronization model. The experimental uncertainty is estimated by assuming the same flavour tagging capabilities than at 250 GeV. Compared to the 250~\gev{} estimate, the impact of systematic uncertainties is much reduced. This is a result of the much larger mass sensitivity at the $Z$-pole, as antipicated in Section~\ref{sec:theory}. 

The expected precision of the \bquark mass measurement is then:
\begin{align}
    \Delta m_{b} (m_Z) = 0.12~\gev = \pm 0.02 (stat.) \pm 0.09 (had.) \pm 0.02 (exp.) \pm 0.06 (th.)~\gev
\end{align}

The total uncertainty of 0.12~\gev, improves considerably on the combination of the LEP/SLD measurements.

\section{Conclusions and summary}

The mass measurements at the ILC can provide an important contribution to the characterization of the scale evolution of the \bquark mass. To visualize their impact, the projections obtained in this section are added to Fig.~\ref{testlabel}. It furthermore includes the world average for the mass measurements at low scale, given by the PDG value for $m_b(m_b)$~\cite{Zyla:2020zbs} and the LEP and SLD measurements~\cite{Abreu:1997ey,Abe:1998kr,Brandenburg:1999nb,Barate:2000ab,Abbiendi:2001tw,Abdallah:2005cv,Bilenky:1998nk,Abdallah:2008ac} of $m_b(m_Z)$.
The reference value of $m_b(m_b) =$ 4.18 GeV is evolved to higher scale using the RGE implemented in the REvolver code~\cite{Hoang:2021fhn}.

The ILC measurements are expected to bring considerable additional evidence for the evolution of the bottom quark mass. The measurement of \Rbl at $\sqrt{s}=$ 250~\gev has a modest precision, limited by the much reduced mass sensitivity, but extends the test of the QCD evolution to energies not probed by previous experiments. The GigaZ program offers the potential of a considerable improvement of the measurement of $m_b(m_Z)$.

\begin{figure}[ht]
\centering
\includegraphics[scale=0.6]{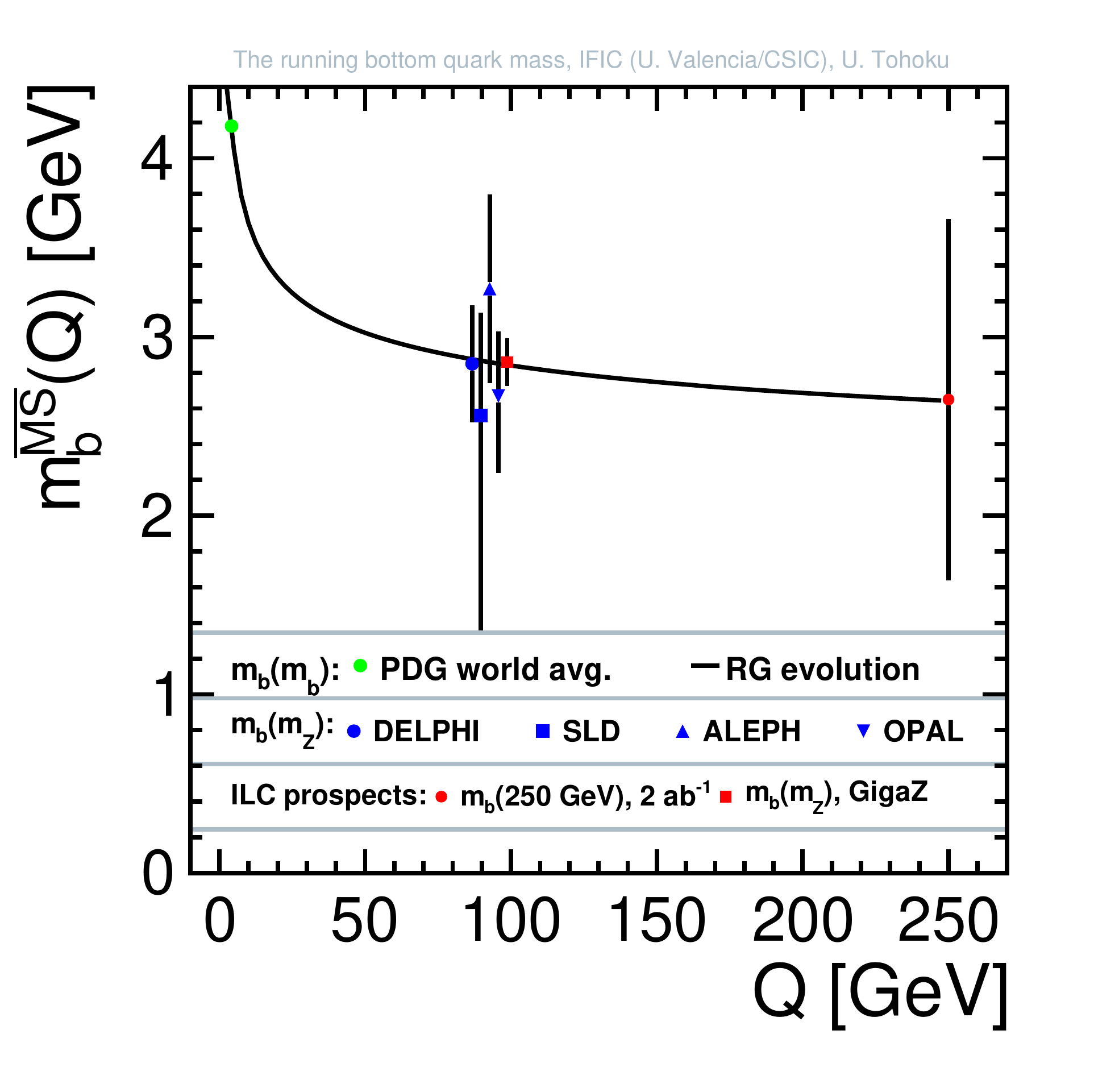}
  \caption{ The evolution of the bottom-quark $\overline{MS}$ mass with the scale $Q$. The black line corresponds to the QCD prediction, obtained from the RGE implemented in the REvolver code~\cite{Hoang:2021fhn} for four loops and input values for $\alpha_s(m_Z)$ and $m_b(m_b)$ given by the PDG world average \cite{Zyla:2020zbs} (see Eqs. \ref{eq:inputmb} and \ref{eq:input}). The reference bottom quark mass is shown in a green marker. The blue markers indicate the results of four measurements at LEP and by SLD in Table~\ref{tab:mbmz}. The red markers correspond to the projections obtained in this note.}
\label{testlabel}
\end{figure}

\section*{Acknowledgements}
We would like to thank the LCC generator working group and the ILD software working group for providing the simulation and reconstruction tools and producing the Monte Carlo samples used in this study.
This work has benefited from computing services provided by the ILC Virtual Organization, supported by the national resource providers of the EGI Federation and the Open Science GRID. 

A. Irles  and M. Vos are  funded by projects FPA2015-65652-C4-3-R (MINECO/FEDER) and PGC2018-094856-B-100, PROMETEO-2018/060 (Generalitat Valenciana), and the iLINK grant (CSIC). A. Irles also acknowledges the financial support from the Generalitat Valenciana (Spain) under the grant number CIDEGENT/2020/21.
G. Rodrigo is supported by the Spanish Government (Agencia Estatal de Investigaci\'on)
and ERDF funds from European Commission (Grant No. FPA2017-84445-P), Generalitat
Valenciana (Grant No. PROMETEO/2017/053), and the COST Action CA16201
PARTICLEFACE.
S. Tairafune acknowledges the financial and educational support by GP-PU (Graduate Program on Physics for the Universe) of Tohoku University. 

\bibliographystyle{JHEP}
\bibliography{references}
\pagebreak
\end{document}